
\input harvmac.tex
\input epsf

\def\figin{\epsfcheck\figin}\def\figins{\epsfcheck\figins}
\def\epsfcheck{\ifx\epsfbox\UnDeFiNeD
\message{(NO epsf.tex, FIGURES WILL BE IGNORED)}
\gdef\figin##1{\vskip2in}\gdef\figins##1{\hskip.5in}
\else\message{(FIGURES WILL BE INCLUDED)}%
\gdef\figin##1{##1}\gdef\figins##1{##1}\fi}
\def\DefWarn#1{}
\def\figinsert{\goodbreak\midinsert}
\def\ifig#1#2#3{\DefWarn#1\xdef#1{fig.~\the\figno}
\writedef{#1\leftbracket fig.\noexpand~\the\figno}%
\figinsert\figin{\centerline{#3}}\medskip\centerline{\vbox{\baselineskip12pt
\advance\hsize by -1truein\noindent\footnotefont{\bf Fig.~\the\figno:} #2}}
\bigskip\endinsert\global\advance\figno by1}

\def\Title#1#2{\rightline{#1}
\ifx\answ\bigans\nopagenumbers\pageno0\vskip0.5in%
\baselineskip 15pt plus 1pt minus 1pt
\else
\pageno1\vskip.25in\fi \centerline{\titlefont #2}\vskip .3in}

\ifx\answ\bigans\def\tcbreak#1{}\else\def\tcbreak#1{\cr&{#1}}\fi

\font\cmss=cmss10 \font\cmsss=cmss10 at 7pt
\def\IZ{\relax\ifmmode\mathchoice
{\hbox{\cmss Z\kern-.4em Z}}{\hbox{\cmss Z\kern-.4em Z}}
{\lower.9pt\hbox{\cmsss Z\kern-.4em Z}}
{\lower1.2pt\hbox{\cmsss Z\kern-.4em Z}}\else{\cmss Z\kern-.4em Z}\fi}
\def\calm{{\cal M}}

\lref\REFmatrixmodels{
D. Gross and A. Migdal,  Nucl. Phys. {\bf B340} (1990) 333\semi
E. Br\'ezin and V. Kazakov, Phys. Lett. {\bf B236} (1990) 144\semi
M. Douglas and S. Shenker, Nucl. Phys. {\bf B335} (1990) 635.}

\lref\REFcequalsone{
D. Gross and N. Miljkovi\'c, Phys. Lett. {\bf B238} (1990) 217\semi 
E. Br\'ezin, V. Kazakov and  A. Zamolodchikov, \hfill\break
Nucl. Phys. {\bf B338} (1990) 673.}

\lref\REFgrossreview{
D. Gross, {\sl The $c=1$ Matrix Models,} in: {\sl Two Dimensional
Quantum Gravity and Random Surfaces} (World Scientific, Singapore, 1992) p.143}

\lref\REFdavid{
F. David, {\sl A Scenario for the  $c>1$ Barrier in
Noncritical Bosonic Strings,} report SACLAY-SPHT-96-112, October 1996,
{\tt hep-th/9610037}.}

\lref\REFEKreduction{
T. Eguchi and  H. Kawai, Phys. Rev. Lett. {\bf 48} (1982) 1063\semi
A. Migdal, Phys. Rep. {\bf 102} (1983) 199.}
 
\lref\REFparisioriginal{
G. Parisi, Phys. Lett. {\bf B238} (1990) 213.}

\lref\REFkostov{
I. Kostov, Nucl. Phys. {\bf B376} (1992) 539.}

\lref\REFparisielse{
G. Parisi, Phys. Lett. {\bf B238} (1990) 209;
Europhys. Lett. {\bf 11} (1990) 595\semi
G. Parisi, G. Salina and A. Vladikas, Phys. Lett. {\bf B256} (1991) 397.}

\lref\REFgrossklebanov{D. Gross and I. Klebanov, 
Nucl. Phys. {\bf B344} (1990) 475\semi
Nucl. Phys. {\bf B354} (1991) 459.}

\lref\REFboulatovkazakov{
D. Boulatov and V. Kazakov, Int. J. Mod. Phys. {\bf A8} (1993) 809\semi
Nucl. Phys. Proc. Suppl. {\bf 25A} (1992) 38.} 

\lref\REFkosterlitzthouless{
V. Berezinskii, JETP {\bf 34} (1972)610\semi
J.M. Kosterlitz and D. Thouless, J. Phys. {\bf C6} (1973) 1181\semi
J. Villain, J. Phys. {\bf C36} (1975) 581.}

\lref\REFgrosswitten{
D. Gross and E. Witten, Phys. Rev. {\bf D21} (1980) 446.}

\lref\REFdouglaskazakov{
M. Douglas and V. Kazakov, Phys. Lett. {\bf B319} (1993) 219.}

\lref\REFbipz{
E. Br\'ezin, C. Itzykson, G. Parisi and  J.B. Zuber,\hfill\break
Commun. Math. Phys. {\bf 59} (1978) 35.}

\lref\REFitzyksonzuber{
C. Itzykson and J.B. Zuber, J. Math. Phys. {\bf 21} (1980) 411.}

\lref\REFhopf{
A. Matytsin, Nucl. Phys. {\bf B411} (1994) 805.}

\lref\REFgrossinstantons{
D. Gross and A. Matytsin, 
Nucl. Phys. {\bf B429} (1994) 50\semi 
J. Minahan and A. Polychronakos,  Nucl. Phys. {\bf B422} (1994) 172.}

\lref\REFwynter{
V. Kazakov, M. Staudacher and T. Wynter,\hfill\break
Commun. Math. Phys. {\bf 177} (1996) 451\semi
Nucl. Phys. {\bf B471} (1996) 309.}

\lref\REFpenner{
Yu. Makeenko, Phys. Lett. {\bf B314} (1993) 197\semi
Int. J. Mod. Phys. {\bf A10} (1995) 2615\semi
L. Paniak and  N. Weiss, J. Math. Phys. {\bf 36} (1995) 2512.}

\lref\REFlittleknown{
D. Boulatov, Mod. Phys. Lett. {\bf A9} (1994) 1963\semi
A. Matytsin and A. Migdal, Int. J. Mod. Phys. {\bf A10} (1995) 421\semi
J.-M. Daul, {\sl $Q$-states Potts Model on a Random Planar Lattice,}
ENS report, November 1994, {\tt hep-th/9502014}.}

\lref\REFgrossinducedQCD{
D. Gross, Phys.Lett. {\bf B293} (1992) 181.}

\lref\REFkazakovmulticritical{
V. Kazakov, Mod. Phys. Lett. {\bf A4} (1989) 2125.}

\lref\REFfirstpaper{
A. Matytsin and P. Zaugg, {\sl Kosterlitz--Thouless
Phase Transitions on Discretized Random Surfaces,} MIT and ENSLAPP report,
November 1996, {\tt hep-th/9611170}.}

\lref\REFmigdalrusakov{
A. Migdal, Sov. Phys. JETP {\bf 42} (1975) 413\semi
Zh. Eksp. Teor. Fiz. {\bf 69} (1975) 810\semi
B. Rusakov, Mod. Phys. Lett. {\bf A5} (1990) 693.}

\lref\REFdouglas{M. Douglas, {\sl Large $N$ Gauge Theory: Expansions and
Transitions,} in: Trieste Spring School 1994, {\tt hep-th/9409098}.}

\lref\REFgrosscylinder{
D. Gross and A. Matytsin, Nucl. Phys. {\bf B437} (1995) 541.}

\Title{\vbox{\baselineskip12pt\hbox{MIT-CTP-2603}
\hbox{ENSLAPP-A-638/97}}}
{\vbox{\centerline{The Two-Dimensional $O(2)$ Model}\vskip0.07in
\centerline{on a Random Planar Lattice}\vskip0.07in
\centerline{at Strong Coupling}}}
\centerline{\vbox{\hsize3in\centerline{Andrei Matytsin}}}
{\it
\smallskip
\centerline{Center for Theoretical Physics}
\centerline{Laboratory for Nuclear Science}
\centerline{Massachusetts Institute of Technology}
\centerline{77 Massachusetts Avenue}
\centerline{Cambridge, MA 02139}}
\smallskip
\centerline{and}
\smallskip
\centerline{
\vbox{\hsize3in\centerline{Philippe Zaugg}}}
%
{\it
\smallskip
\centerline{Laboratoire de Physique Th\'eorique
ENSLAPP\footnote
{$^*$}{URA 14-36 du CNRS, associ\'ee \`a l'Ecole Normale Sup\'erieure de
Lyon et \`a l'Universit\'e de Savoie.}}
\centerline{LAPP, Chemin de Bellevue, BP 110}
\centerline{F-74941 Annecy-le-Vieux Cedex, France\footnote{}{e-mail:
{\tt matytsin@marie.mit.edu, zaugg@lapp.in2p3.fr}}
}}
%
\bigskip
\centerline{\bf Abstract}
\smallskip
\noindent
The large spacing phase of the infinite random matrix chain, which represents
the strongly coupled two-dimensional $O(2)$ model on a random planar lattice,
is explored. A class of solutions 
valid for large lattice spacings 
is constructed. It is proved that these 
solutions exhibit the critical exponents characteristic of pure 
two-dimensional
gravity. 
The character expansion for the chain model is developed and 
an order parameter governing the Kosterlitz--Thouless phase transition
is identified.
%

\secno 0
\newsec{Introduction}

String theories with the matter central charge one are of special interest.
They are positioned in between the exactly solved $c<1$ minimal models 
interacting with
quantum gravity \REFmatrixmodels\ 
and the yet mysterious $c>1$ noncritical string 
theories \REFgrossreview\REFdavid.

The $c=1$ string where the target space is an infinite one-dimensional 
line can   
also be analyzed exactly \REFcequalsone. 
However, new problems arise once the
target space is compactified. In that case the theory has been solved only 
for sufficiently large values of the compactification radius $R>R_{\rm cr}.$
At $R=R_{\rm cr}$ the $c=1$ model undergoes a phase transition, and there
is evidence that for smaller $R$ it behaves like a theory of pure 
two-dimensional gravity, without any matter at all 
\REFgrossklebanov\REFboulatovkazakov.

The transition at $R=R_{\rm cr}$ is quite similar in nature to the famous 
Kosterlitz--Thouless phase transition in the two-dimensional $O(2)$ 
model \REFkosterlitzthouless.
Naively, these transitions are not automatically present in continuum theory
and  appear only if the worldsheet is discretized. Both of them are induced
by topologically nontrivial field configurations---vortices---where
the compactified string, or a two-component unit vector in the $O(2)$ model
winds around a unit circle as we follow the boundary of an elementary 
worldsheet plaquette. 
The dynamical role of vortices in the theory depends on how the
vortex energy and entropy compare to each other.
As a result, while the effect of vortices is negligible for 
$R>R_{\rm cr}$ (or in the weakly coupled phase of the $O(2)$ model)
the vortices actually dominate the dynamics at $R<R_{\rm cr}$ (respectively,
in the strong coupling phase of the $O(2)$ model.) 
Furthermore, the well known $R\leftrightarrow\alpha^{\prime}/R$ duality of 
one-dimensional compactified string theory holds only if the vortices 
are completely ignored. It is possible to show \REFgrossklebanov\
that the contributions to the free energy coming from the 
vortexless sector of the $c=1$ theory do respect the $T$-duality, whereas
the vortex contributions manifestly break it.

To understand the dynamics of vortices interacting with quantum gravity is
a longstanding problem \REFgrossklebanov\REFboulatovkazakov. 
Technically, this problem can be formulated in a very natural way 
using the language of random matrix models.
Indeed, there the two-dimensional worldsheet is discretized
by construction. Thus any matrix model describing the compactified 
one-dimensional string theory will contain vortices from the very beginning. 
The difficulty is, all such models involve an infinite number of interacting 
matrices and, because of that, have not yet been solved.

In this paper we shall explore the simplest among these models,
a one-dimensional infinite random matrix chain 
defined by the partition 
function \REFgrossklebanov\REFparisioriginal\REFparisielse\REFkostov\
\eqn\chain{
{\cal Z}=\int\prod\limits_{n=-\infty}^{+\infty}\!\! dM_n\thinspace
\exp\biggl\{ -N\thinspace {\rm Tr}\sum_{n=-\infty}^{+\infty} 
\biggl[{(M_{n+1}-M_n)^2\over 2\epsilon} + \epsilon V(M_n)\biggr]
\biggr\}.}
Here all $M_n$ are $N\times N$ Hermitian matrices while $V(M)$
is a polynomial potential such as, for example, $m^2 M^2/2+{\tilde g} M^3/3$ or
$m^2 M^2/2 + {\tilde g} M^4/4$.

It is possible to show \REFgrossklebanov\
that the large $N$ limit of the chain model
does indeed describe  the one-dimensional bosonic string theory compactified
on a circle of radius $R=1/\epsilon$ or, equivalently, 
the two-dimensional 
$O(2)$ nonlinear sigma model coupled to quantum gravity.
That is to say, the leading term in the large $N$ expansion 
of ${\cal Z}$ summarizes correctly the vortex properties on topologically
spherical random surfaces.
At the same time the $1/N$ corrections to the chain partition function 
do not represent the higher genus compactified $c=1$ amplitudes.
Instead, they correspond to a different theory---a one-dimensional
string with discretized target space. On a spherical worldsheet
this theory is related to the compactified one-dimensional string by 
a duality transformation, and the string partition functions
for the two coincide.

It has long been known \REFgrossklebanov\
that the infinite matrix chain undergoes a 
Kosterlitz--Thouless phase transition at a certain 
$\epsilon=\epsilon_{\rm cr}.$ Throughout the region 
$\epsilon<\epsilon_{\rm cr}$ where vortices can be neglected
the critical indices of \chain\ do correspond to $c=1$. On the other hand, 
very little, if anything, is known about the properties of the matrix chain 
for $\epsilon>\epsilon_{\rm cr}.$ Quite remarkably, it is this regime, totally
dominated by vortices, that is the hardest to analyze by traditional 
means.

Certainly, in the limit of infinitely large $\epsilon$ the matrix chain 
partition function decouples into a product of independent one-matrix models.
Therefore, one might conjecture that, at least for large $\epsilon$, the 
critical indices of ${\cal Z}$ are those of pure two-dimensional gravity.
We shall see that such  a conclusion is in fact true 
for all $\epsilon>\epsilon_{\rm cr}.$ As a consequence, the effects of 
dimensional reduction in string theories and field theories may in fact 
be much 
more similar than one usually believes. Consider, say, a field theory on a 
manifold with a compact dimension. When the compactification radius vanishes
the compactified dimension effectively disappears, and we obtain a 
(perhaps modified) field theory in one less dimension. In string theory the
picture is quite different. There because of the 
$R\leftrightarrow\alpha^{\prime}/R$ duality the small and large 
compactification
radii appear to be strictly equivalent, and taking $R$ to zero does not reduce
the effective dimension.
However, once the nonperturbative effects---such as vortices---are included,
the situation may change drastically and the parallels with field theory may 
get restored. Exactly that happens in our 
one-dimensional example
where for $R<R_{\rm cr}$ the compactified string theory behaves as if the
one-dimensional target space circle had completely disappeared
and the target space dimension was equal to zero.

Below we shall
construct, for certain $V(M)$, an exact solution of the matrix chain
valid at  $\epsilon>\epsilon_{\rm cr}.$
Remarkably, the computational tools needed for that are quite interesting on 
their own. As we shall see, the large $N$ asymptotics of the matrix chain 
model is related to a certain one-dimensional hydrodynamic system. The 
Kosterlitz--Thouless phase transition corresponds then to the formation of 
a shock-type configuration in the moving fluid.

To state this more precisely,
let us first rescale all matrices $M_n=\sqrt{\epsilon}{\cal M}_n$ so that
$\epsilon$ disappears from the kinetic term in the partition function,
\eqn\chaintwo{
{\cal Z}=\int\prod\limits_{n=-\infty}^{\infty}\!\! d\calm_n
\; \exp\biggl\{N \, {\rm tr}\!\!\sum_{n=-\infty}^{\infty}
\Bigl[ \calm_n\calm_{n+1} - U(\calm_n)\Bigr]\biggr\}}
the new potential $U(\calm)$ being related to $V(M)$ by
\eqn\potentials{
U(\calm)=\calm^2 +\epsilon V(\sqrt{\epsilon} \calm).}
At $N\to \infty$ the eigenvalues
$\lambda_{1,n}\dots \lambda_{N,n}$ of any given matrix ${\cal M}_n$ 
in \chaintwo\
condense to form a smooth distribution $\rho_n(\lambda).$ Actually, due to
the translational invariance of the matrix chain this distribution is the same
for any chain site, $\rho_n(\lambda)\equiv \rho(\lambda)$. It turns out that
the density $\rho(\lambda)$ can be found as a solution to the following 
hydrodynamic problem \REFhopf\REFfirstpaper. 
Consider a one-dimensional droplet of  compressible 
fluid with a special equation of state 
\eqn\eqnstate{P=-{\pi^2\over 3} \rho^3}
where $P$ is the local pressure and $\rho$ ---the local fluid density.
Imagine that at time $t=0$ the density $\rho$ equals (the unknown) $\rho(x)$
and the initial fluid velocity is
\eqn\initvelocity{
v(x)=
{1\over 2}U^{\prime}(x)-x.}
Now, demand that after one unit of time, at $t=1$, the density of the droplet 
returns to its initial value,
\eqn\returninitial{
\rho(x, t=1)=\rho(x, t=0)=\rho(x).}
This condition fixes $\rho(x)$ once $U(x)$ (and hence $v(x)$) is given, so that
the problem of the matrix chain reduces to solving the hydrodynamic Euler 
equations.

As a consequence of \returninitial, together with the time reversal 
symmetry of Euler equations, the initial and final fluid velocities 
are opposite, $v(x, t=1)=-v(x, t=0)$ and, furthermore, $v(x, t=1/2)\equiv 0$.
The density at $t=1/2$, which we denote $\rho_{1/2}(x)$, shall play a very 
important role in our analysis. Its value at the origin $\zeta=\rho_{1/2}(0)$
provides an ``order parameter'' for the Kosterlitz--Thouless transition.
That is to say, $\zeta\equiv 0$ for any $\epsilon<\epsilon_{\rm cr}$ but
$\zeta>0$ everywhere in the phase of large $\epsilon$.

Of course, the Kosterlitz--Thouless transition is not explicitly associated 
with the spontaneous breaking of any symmetry and using the term ``order
parameter'' may appear purely formal. However, this phase transition
does reflect a nonanalytic change in the group representation structure of 
\chain\ at large $N$. In section 4 we shall demonstrate that the integrand of 
${\cal Z}$ can be expanded as a series over characters of $SU(N)$. At 
infinite $N$ only one representation in this series, with the 
Young tableau row lengths
$(n_1, \dots, n_N)$ all of order $N$, makes an important contribution.
Such representations are usually characterized by the Young tableau 
density $\rho_l(h)$ defined as the density of points $h_i=(n_i-i)/N+1$
in a small interval around $h$. Since the row lenghts are all ordered, 
$\rho_l(h)$ can never be greater than one.

\ifig\youngtableausfigure{
The Young tableaus for the representations dominant at $\epsilon<1$ (left)
and $\epsilon>1$ (right).}
{\epsfxsize2.0in\epsfbox{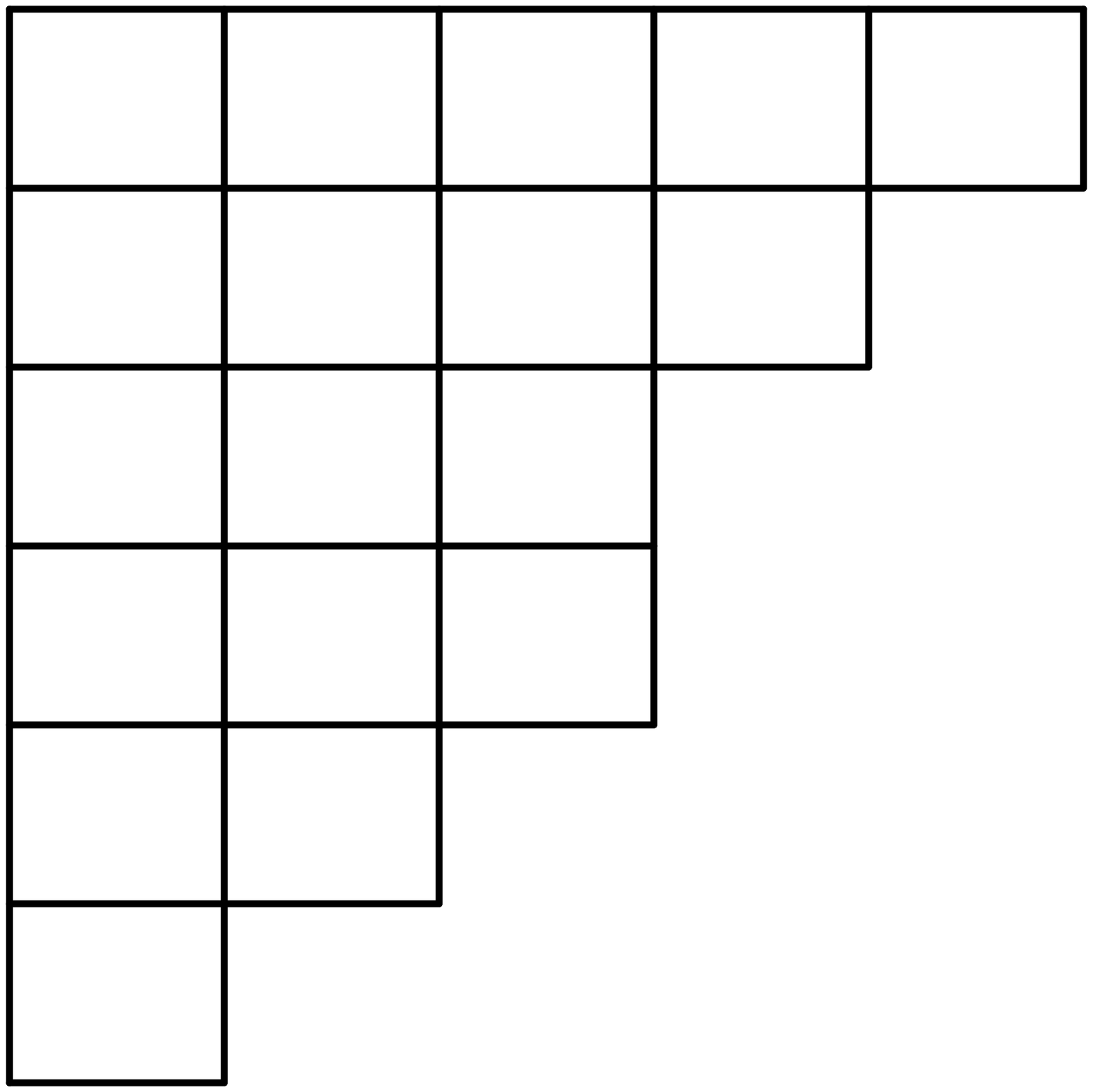}
\hskip0.5in\epsfxsize2.0in\epsfbox{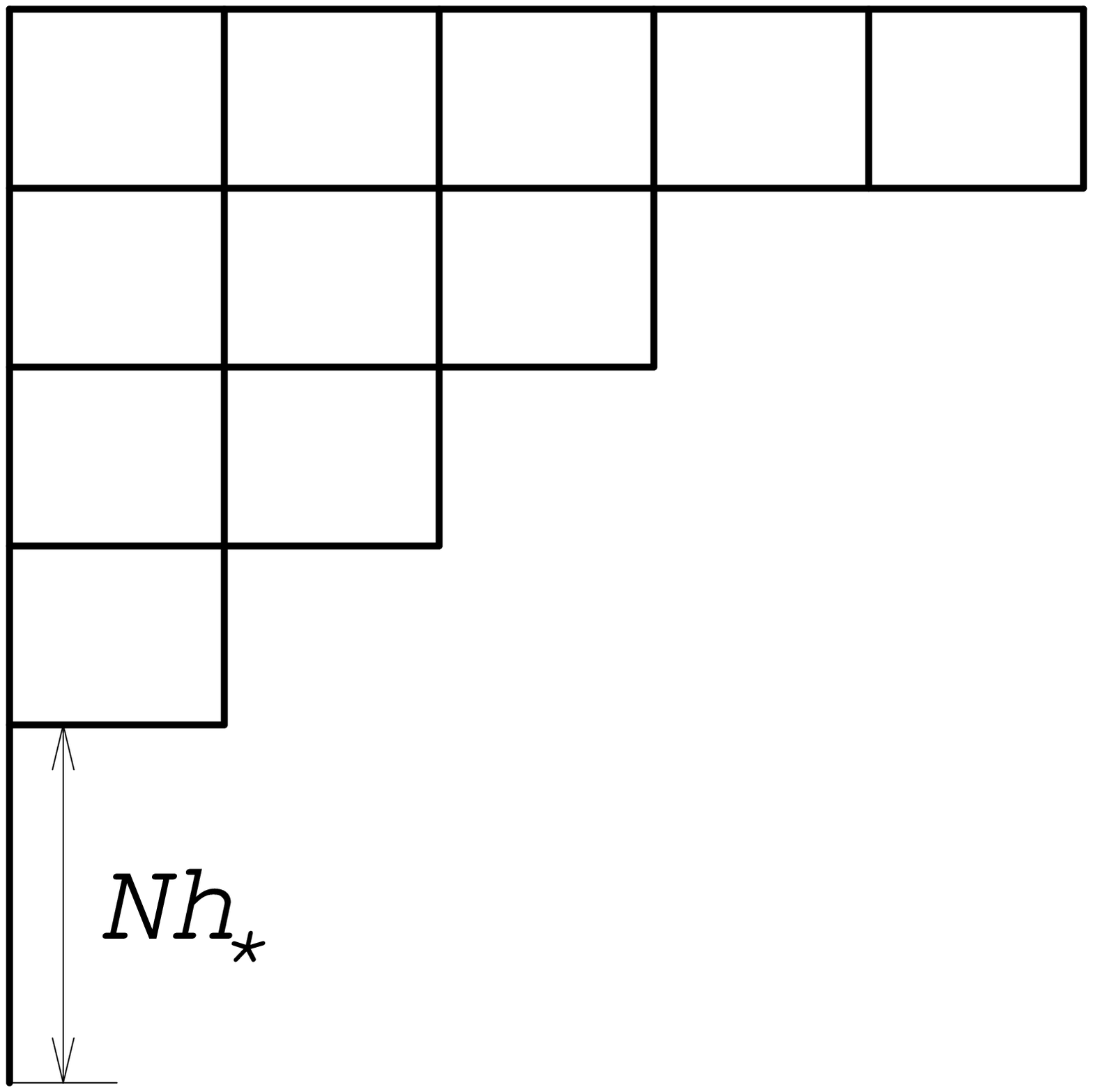}}

Whenever the Young tableau density saturates this upper bound, $\rho_l(h_*)=1$,
the corresponding Young tableau develops a gap. It turns out that such gap
formation is just another aspect of the same Kosterlitz--Thouless phase
transition at $\epsilon=\epsilon_{\rm cr}$. Moreover, the ``order
parameter'' $\zeta=\rho_{1/2}(0)$ happens to be very directly related to 
the gap size $h_*$,
\eqn\zetaeqn{
\zeta=\rho_{1/2}(0)={2\over \pi} \sqrt{h_*}.}
Such a correspondence unifies, at least technically, the 
Kosterlitz--Thouless transition and the Douglas--Kazakov phase transitions
known from two-dimensional QCD and the dually weighted graph 
models \REFdouglaskazakov\REFwynter.
The analogy between those transitions goes, in fact, rather far.
Dynamically, all of them are induced by the topologically  
nontrivial classical solutions of underlying theories---vortices in the
$O(2)$ model and instantons in QCD$_2$ respectively, which become statistically
dominant at strong coupling \REFgrossinstantons.

Below we shall first explore the large $\epsilon$ limit of the chain model
\chain. This shall be done in the next section. In section 3 we shall 
construct a family of exact solutions describing the matrix chain in the 
interval $\epsilon>\epsilon_{\rm cr}$. The character expansion methods
shall be presented in section 4. Finally, in section 5 we shall confirm
our conclusions by studying an explicit $1/\epsilon$ expansion for the case 
of a pure quartic potential $U(x)$.

\newsec{The Infinite Matrix Chain at Large Lattice Spacings}

To begin with, we shall demonstrate how the hydrodynamic representation 
\eqnstate--\returninitial\ reproduces the solution of the infinite 
matrix chain 
in the limit $\epsilon\gg \epsilon_{\rm cr}$. 
We shall use this special case to understand certain properties of the 
model---especially the analytic structure of the midway eigenvalue density
$\rho_{1/2}(x)$ ---which turn out to hold more generally for any finite
$\epsilon>\epsilon_{\rm cr}$.

For large 
$\epsilon$ the ``kinetic'' term in the matrix model action,
${\rm tr}(M_{n+1}-M_n)^2/2\epsilon$ is much smaller than the 
potential energy term $\epsilon \, {\rm tr}V(M_n)$. 
Equivalently, one can neglect the nearest neighbor 
coupling term ${\rm tr}({\cal M}_n{\cal M}_{n+1})$ in \chaintwo\
thereby reducing ${\cal Z}$ to a product
of identical one-matrix partition functions
%
\eqn\factorization{
{\cal Z}=\prod\limits_{n=-\infty}^{+\infty}\int d{\cal M}_n\, \exp\big[
-N\, {\rm tr}U({\cal M}_n)\big].}
The eigenvalue density $\rho(\lambda)$ for the factorized 
ensemble \factorization\
is the same as the density in a one-matrix model with the potential 
$U({\cal M})$ and can be found from the Riemann--Hilbert equation \REFbipz\
\eqn\riemannhilbert{
-\kern-1.1em \int {\rho(\lambda^{\prime})\, d\lambda^{\prime}\over
\lambda-\lambda^{\prime}}={1\over 2} U^{\prime}(\lambda).}
For concreteness let us consider the specific case of a quartic double-well
potential
\eqn\doublewell{
V(M)={m^2\over 2} M^2+ {{\tilde g}\over 4}M^4}
with $m^2<0$ and ${\tilde g}>0$. By rescaling $M, {\tilde g}$ and the
lattice spacing $\epsilon$ it is always possible to make the absolute
value of $m^2$ anything we want. From now on we shall choose $m^2=-4$.
As we shall see later, for this $m^2$ the position of the 
Kosterlitz--Thouless point is fixed at $\epsilon_{\rm cr}=1$.
The potential $U$ corresponding to such $V$ has the form 
\eqn\pott{
U(x)=-{\mu^2 x^2\over 2} +{g x^4\over 4}}
with $\mu^2=2(2\epsilon^2-1)$ and $g={\tilde g} \epsilon^3$.
%

Strictly speaking, the one-matrix model with a double-well $V(M)$
describes not the $c=0$ but $c=-2$ quantum gravity. To represent
the pure two-dimensional quantum gravity one must change 
the overall sign of $V$ taking $m^2>0$ and ${\tilde g}<0$.
The computational methods presented below are equally applicable in both
cases but the formulas happen to be simple for the double-well model.
Therefore we shall discuss the double-well $c=-2$ theory first and then
outline the modifications arising when $c=0$.

For the potential $U(M)$ given by \pott\ the solution of the 
Riemann--Hilbert equation \riemannhilbert\ is quite easy to find directly.
At sufficiently large ${g}$ (that is, in the strong coupling phase)
such solution is given by 
\eqn\onematstrong{
\pi\rho(x)=(Ax^2+B)\sqrt{1-C^2 x^2}}
where the coefficients $A$, $B$ and $C$ equal
\eqn\coefftstr{\left\{
\eqalign{
&C^2={1\over 8}\bigl(\sqrt{\mu^4+12 g}- \mu^2\bigr)\cr
&B={1\over 6C}\bigl(\sqrt{\mu^4+12 g}- 2\mu^2\bigr)\cr
&A={g\over 2C}.\cr}\right.}
%
Equation \onematstrong\ predicts the eigenvalue density at $x=0$ to be 
$\rho(0)=B/\pi$, thereby implying $B>0$. This condition is satisfied only for
$g>g_{\rm cr}=\mu^4/4$.
When the coupling $g$ decreases below $\mu^4/4$ the
one-matrix model undergoes a phase transition and the  
eigenvalue density $\rho(x)$ develops 
a two-cut structure, 
\eqn\onematweak{
\pi\rho(x)={gx\over 2}\sqrt{(x^2-b^2)(a^2-x^2)}}
with
\eqn\coefftweak{
\left\{\eqalign{
a^2&={1\over g}(\mu^2+2\sqrt{g})\cr
b^2&={1\over g}(\mu^2-2\sqrt{g}).\cr}\right.}
At the critical point, where 
$g=g_{\rm cr}=\mu^4/4$ the solutions for the strong and weak couplings 
coincide. They yield the eigenvalue density which vanishes at $x=0$,
\eqn\rhocrit{
\pi\rho_{\rm cr}(x)={\mu^3 x^2\over 2\sqrt{2}}\sqrt{1-{\mu^2 x^2\over 8}}.}
For small $x$ this critical density behaves as $\rho_{\rm cr}(x)
\propto |x|^{\delta}$ with $\delta=2$. The value of the 
critical index $\delta$ is  a universal property of the one-matrix model---it
does not change even when 
we modify the matrix model potential by any 
generic higher
order terms. 
In fact, $\delta$ is related to the so-called string susceptibility
$\gamma_{\rm str}=1-\delta$, a quantity which plays an important role in 
string theory.

Let us see how the critical density, and hence the universal property
$\rho_{\rm cr}(x)\propto x^2$, can be extracted from the hydrodynamic picture.
The Euler equations which govern the motion of the one-dimensional
fluid are
\eqn\euler{\left\{\eqalign{
&{\partial \rho\over \partial t}+{\partial \over \partial x}[\rho v]=0\cr
&{\partial v\over \partial t}+v {\partial v\over \partial x}
=-{1\over \rho}{\partial P\over \partial x}\cr}\right.}
where  $P$ is the fluid pressure, 
$\rho$ ---its density and $v$ ---the velocity.
For the special case when $P$ and $\rho$ are related by \eqnstate\
these equations can be simplified and even integrated.
To this end, introduce the complex valued function $f=v+i\pi\rho.$
In terms of $f$ the system \euler\ reduces to the complex Hopf equation
\eqn\hopf{
{\partial f\over \partial t}+ f{\partial f\over \partial x}=0}
which has the following formal solution
\eqn\implicit{
f(x, t)=f_0\big[x-t f(x,t)\big].}
The function $f_0(x)=f(x, t=0)$ represents the requisite initial data 
for the Hopf equation.
In our problem, one specifies only the real part of $f_0$ given by 
\initvelocity\ while the imaginary part ${\rm Im}\, f_0(x, 0)=\pi\rho(x)$
must be calculated to satisfy the boundary conditions \returninitial.

Due to the time reversal symmetry of Euler equations, imposing 
\returninitial\ is equivalent to demanding that the velocity 
at $t=1/2$ vanishes identically, $v(x, t=1/2)\equiv 0$.
Therefore, the value of the Hopf function $f(x, t)$ at $t=1/2$
ought to be purely imaginary, $f(x, 1/2)=i\pi\rho_{1/2}(x)$.
As a result, the complex equation 
\eqn\implicithalf{
i\pi\rho_{1/2}(x)=f_0\bigl[x-\half i\pi\rho_{1/2}(x)\bigr]}
must have a real-valued solution for $\rho_{1/2}(x)$.

In the limit of large $\epsilon$ the critical density \rhocrit\ and, more
generally, \onematstrong\ or \onematweak, do give rise to a real $\rho_{1/2}$.
For example, if $\rho(x)=\rho_{\rm cr}(x)$ the corresponding initial
Hopf function equals
\eqn\fzerocrit{
\eqalign{
f_0(z)&=v(z)+i\pi\rho_{\rm cr}(z)\cr
&=-z-{\mu^2\over 2} z + {\mu^4\over 8}z^3 - {\mu^4 z^3\over 8}\sqrt{1-{8\over 
\mu^2 z^2}}\cr}}
where we have used \initvelocity, substituted $g=g_{\rm cr}=\mu^4/4$
and continued $f_0(z)$ to $z>2\sqrt{2}/\mu$.
Now, we shall see in a moment that the typical values of 
$z=x-i\pi\rho_{1/2}(x)/2$ appearing in equation \implicithalf\ are of 
order one, $z\sim{\cal O}(1)\gg1/\mu$.
Therefore, we can expand $f_0(z)$ for large $\mu$ and finite $z$, 
\eqn\fzeroapprox{
f_0(z)=-z+{1\over z}+{\cal O}(\mu^{-2})}
which, together with \implicithalf\ implies
\eqn\rhohalfapprox{
\pi\rho_{1/2}(x)=2\sqrt{1-x^2} +{\cal O}(\mu^{-2}).}
A careful inspection of the above computation shows that equation
\rhohalfapprox\ holds for any eigenvalue density satisfying
the Riemann--Hilbert equation \riemannhilbert.
Actually, it is simply a consequence of the normalization condition
$\int\rho(x)\, dx=1$. Indeed, it is easy to compute $f_0(z)$ given any general
$U(x)$,
\eqn\fzerogeneral{
\eqalign{
f_0(z)&={1\over 2} U^{\prime}(z)-z +i\pi\rho(z)\cr
&=-z+\int {\rho(y)\, dy\over z-y}.\cr}}
For an eigenvalue distribution localized around zero in a small interval
of width $\propto1/\mu$ this formula can be expanded in $1/z$ to yield
\fzeroapprox\ and, consequently, \rhohalfapprox.

As a result, the shape of $\pi\rho_{1/2}(x)$ in the large $\mu$ approximation
says very little about the properties of the underlying 
matrix model. In fact, if we started with $\pi\rho_{1/2}(x)=2\sqrt{1-x^2}$
at $t=1/2$ and
let it evolve until $t=1$ according to the Euler equations, that distribution
would have collapsed to a delta function. The actual properties of
the matrix chain are therefore encoded not in the semicircular 
shape of \rhohalfapprox\ but rather in tiny perturbations over it. These 
perturbations are strongly amplified as the distribution collapses and produce 
the nontrivial eigenvalue density $\rho(x, t=1)=\rho(x)$ at the final
point of time evolution.

Quite remarkably, the effect of small perturbations becomes very transparent
if we consider the analytic structure of $\rho_{1/2}(x)$. One discovers that
$\rho_{1/2}$ has, in general, several different regular branches.
Equation \rhohalfapprox\ approximates, of course, only one of them---the
``physical'' branch. However, for $\epsilon>\epsilon_{\rm cr}$
it happens to be another ``hidden'' branch that actually controls the shape of 
$\rho(x)\equiv\rho(x, t=1)$. To see this take $f_{1/2}(x)=i\pi\rho_{1/2}(x)$
as the initial condition for the Hopf equation and evolve it 
backwards in time thus obtaining 
$f(x, t=0)=f_0(x)$.  The functions $f_{1/2}$ and $f_0$ are related by a 
formula analogous to \implicithalf
\eqn\implicitforward{
f_0(x)=f_{1/2}\big[x+\half f_0(x)\big].}
Now we can substitute $x=-iy$ and use 
the explicit form of $f_0(x)$ from \fzerocrit\ to
derive the following parametric representation for $\pi\rho_{1/2}$
\eqn\parametric{
\left\{\eqalign{
&\pi\rho_{1/2}(i\xi)=F(y)\cr
&\xi=-y+{1\over 2}F(y)\cr}\right.}
where
\eqn\Flarge{
F(y)=y+{\mu^2\over 2} y +{\mu^4\over 8}y^3 - {\mu^4 \over 8}y^2\sqrt{y^2+
{8\over \mu^2}}.}
Let us emphasize that this representation applies to the special (but
most interesting) case when the matrix model coupling $g$ has been tuned
to the critical value $g=g_{\rm cr}(\epsilon)$. 

The plot of $\pi\rho_{1/2}(i\xi)$ as a function of $\xi$ is shown in fig.~2
and clearly reveals two distinct branches.  For small $\xi$ one of them 
behaves as $\pi\rho_{1/2}(i\xi)\approx 2\sqrt{1+\xi^2}$ which is simply an
anaytic continuation of the physical branch \rhohalfapprox.
The values of the parameter $y$ corresponding to this branch are of order one, 
$y\sim 1\gg1/\mu$, so that $F(y)=y+1/y+{\cal O}(\mu^{-2}).$ 

Another branch of $\pi\rho_{1/2}$ ---call it the ``hidden'' branch---is 
obtained when $y\ll1/\mu$. Then for large $\mu$ we have
$F(y)\approx y(1+\mu^2/2)$ and therefore
\eqn\anotherbranch{
\pi\rho_{1/2}(i\xi)=2\xi\biggl(1+{4\over \mu^2}\biggr)+\dots.}
To demonstrate the importance of this hidden branch 
let us imagine that we have 
expanded the critical eigenvalue density \rhocrit\ in powers of $x$. 
As it turns out, such an expansion is in one-to-one correspondence 
with the expansion of the hidden branch in powers of 
$\xi$. That is to say, given a power series  for $\pi\rho_{1/2}(i\xi)$ in
\anotherbranch\ and using equation \implicitforward\ one can restore, order
by order, the power series for $\rho_{\rm cr}(x)$.

\ifig\twobranchesfigure{
The midway density $\pi\rho_{1/2}(i\xi)$ for imaginary arguments $x=i\xi$,
and the graphic solution of equation $\pi\rho_{1/2}(i\xi)=\xi/\tau$.
The plot clearly shows two branches of $\pi\rho_{1/2}$.}
{\epsfysize 2.0in\epsfbox{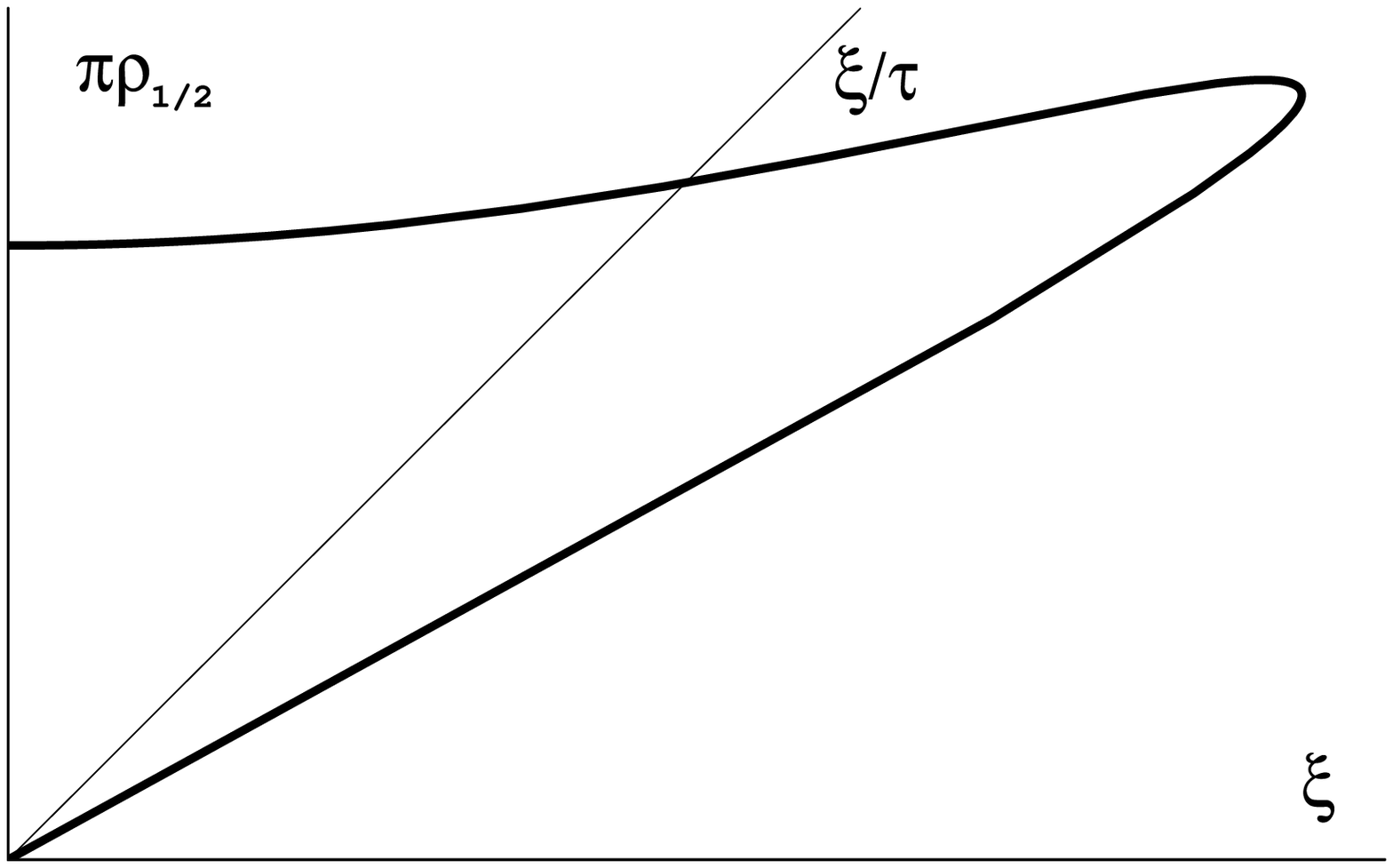}}

The same would not by any means be true for the semicircular physical 
branch \rhohalfapprox. There an expansion around zero has nothing to do with
how $\rho_{\rm cr}(x)$ behaves at small $x$. From this point of view, it is
quite satisfactory that the complicated and nonlocal perturbations 
of the physical branch can be summarized in a simple Taylor
expansion of another 
branch.

Of course, to get the particular graph shown in fig.~2 we have used 
the one-matrix model limit $\epsilon\gg 1$. However, we shall see in the 
next section that the plot of $\pi\rho_{1/2}(i\xi)$ remains qualitatively 
the same for any $\epsilon>1$. Therefore, the conclusions derived from such a 
plot---say, the existence of two branches for $\pi\rho_{1/2}(i\xi)$\ ---are not
at all restricted to the large $\epsilon$ limit.

For a finite $\epsilon$ the hidden branch of $\pi\rho_{1/2}(i\xi)$
is no longer approximated by \anotherbranch. Instead, it has a certain---yet
unknown---Taylor expansion
\eqn\taylor{
\pi\rho_{1/2}(i\xi)=a_1(\epsilon)\xi +a_2(\epsilon)\xi^2+a_3(\epsilon)\xi^3+
\dots}
all the Taylor coefficients being functions of $\epsilon$.
As we advertized above, this expansion can be used, together with
\implicitforward, to reconstruct the potential $U(x)$ and the
eigenvalue density $\rho(x)=\rho(x, t=0)$.
To this effect we consider the expansion 
\eqn\fexpansion{
f_0(x)=f_1 x+ f_2 x^2 + f_3 x^3+\dots}
and solve \implicitforward\ order by order in powers of $x$.
This yields
\eqn\fcoefficients{
\left\{\eqalign{
f_1&= - {2 a_1(\epsilon)\over a_1(\epsilon)-2}\cr
f_2&=i {8  a_2(\epsilon)\over \big[a_1(\epsilon)-2\big]^3}\cr
f_3&= {32a_2^2(\epsilon)\over
\big[a_1(\epsilon)-2\big]^5}
-{16 a_3(\epsilon)\over
\big[a_1(\epsilon)-2\big]^4}
\cr}\right.}
and so forth. The real part of $f_0(x)$ is related, by virtue of 
\initvelocity\ and \pott, to the parameters $\mu^2$ and $g$. In particular,
$f_1=-(\mu^2/2+1)=-2\epsilon^2$ which fixes $a_1(\epsilon)$ in terms of
$\epsilon$
\eqn\aone{
a_1(\epsilon)={2\epsilon^2\over \epsilon^2-1}.}
We see that the hidden branch develops a singularity and ceases to exist
at $\epsilon=\epsilon_{\rm cr}=1$. This is precisely the location of the 
Kosterlitz--Thouless phase transition in the infinite matrix chain with the 
potential \pott. Remarkably, the same critical value of $\epsilon_{\rm cr}$
can be derived independently by studying the singularities
of the $\epsilon<1$ phase and by a variety of other 
methods \REFgrossklebanov\REFboulatovkazakov\REFfirstpaper.

Therefore, the two-branch structure is a characteristic 
property distinguishing the $\epsilon>1$ phase of the infinite 
matrix chain. It contrasts sharply with the 
one-branch behavior found in the $\epsilon<1$ phase of the same 
theory \REFfirstpaper.
Indeed, for $\epsilon<1$ the small $x$ properties of the critical
eigenvalue density can be restored from the
physical $\pi\rho_{1/2}(x)$ alone, without continuing it to any other branches.

The presence of the second branch has a very interesting reflection in how
the droplet evolves with time. It results in the formation of a shock---a
state where the spatial derivative of the fluid density becomes infinite
at a certain point. To see how this happens
consider the time dependence of the 
droplet density at the origin $\rho(x=0, t)\equiv d(t)/\pi$. 
We can determine the function $d(t)$ from an implicit relation analogous
to \implicit\ 
\eqn\implicithalfback{
f(x, t)=f_{1/2}\big[x+\tau f(x, t)\big]}
$\tau$ being the shifted time $\tau=1/2-t$.
Setting here $x=0$ and remembering that $v(x=0, t)=0$ at any $t$, we 
obtain a transcendental equation on $d(t)$
\eqn\tansc{
d(t)=\pi\rho_{1/2}\big[i\tau d(t)\big].}
This equation can be easily solved graphically using the plot of $\pi\rho_{1/2}
(i\xi)$ shown in fig.~2. One simply writes $\xi=\tau d(t)$ so that
\eqn\graphic{
\pi\rho_{1/2}(i\xi)={\xi\over \tau}}
and finds the intersection point of the plot for $\pi\rho_{1/2}(i\xi)$ 
with the straight line of the slope $1/\tau$. At $t=1/2$, when $\tau=0$,
this yields a nonzero $d$ given by the value of the physical branch at $x=0$. 
As we go backwards in time away from $t=1/2$, the fluid
density at the origin changes. In the specific case depicted in fig.~2
it first grows and then starts decreasing until, at a certain moment
$\tau=\tau_{\rm cr}$, the density at $x=0$ vanishes. This critical time
$\tau_{\rm cr}$ is determined by the slope of the hidden branch at the 
origin
\eqn\taucrit{
{1\over \tau_{\rm cr}}= {d\, \pi\rho_{1/2}(i\xi)\over d\xi}\biggr|_{\xi=0}.}
For any $\tau>\tau_{\rm cr}$ or, equivalently, for $t<t_{\rm cr}=1/2-
\tau_{\rm cr}$ equation \graphic\ admits only the solution
$d(t)=0$. Note that, as a consequence of \aone, the critical time
is a function of the lattice spacing $t_{\rm cr}=1/(2\epsilon^2)$.

As a result, the droplet density at the origin remains strictly zero 
at $0\le t\le t_{\rm cr}$ and becomes positive in the interval 
$t_{\rm cr}<t\le 1/2$. Of course, the densities at larger times 
$t\in[1/2, 1]$ are
easily obtained using  the time reflection symmetry $d(t)=d(1-t)$.
The shock-type configuration forms at $t=t_{\rm cr}$ and $t=1-t_{\rm cr}$.
This can be understood by solving equation \implicithalfback\
for small, but nonzero, values of $x$ and times $t$ close to $t_{\rm cr}$.
At such $t$ the graphical solution to equation \graphic\ yields
a value of $\xi$ very close to zero. Consequently, we can use the Taylor 
expansion of $\pi\rho_{1/2}(i\xi)$, given by \taylor, as input for equation
\implicithalfback\ thereby obtaining
\eqn\smallx{
\rho(x, t)\propto\left\{\matrix{
& {\displaystyle a_2(\epsilon)\biggl({\tau_{\rm cr}
\over \tau-\tau_{\rm cr}}\biggr)^3 x^2}
\hfill&{\rm when\ } t<t_{\rm cr}\hfill\cr
& {\displaystyle \sqrt{{a_1^3(\epsilon) \over 2 a_2(\epsilon)}x}}
\hfill&{\rm when\ } t=t_{\rm cr}\hfill\cr
& {\displaystyle 
{\tau_{\rm cr}-\tau \over a_2(\epsilon) \tau^2  \tau_{\rm cr}}
+a_2(\epsilon) \biggl({\tau_{\rm cr}\over \tau_{\rm cr}-\tau}\biggr)^3 x^2}
\hfill& {\rm when\ } t>t_{\rm cr}\hfill\cr}\right.}
for $x\to 0$ and $|t-t_{\rm cr}|\ll 1$.
A typical plot of $\rho(x, t)$ at various times is shown in fig.~3.
When $t<t_{\rm cr}$ the fluid flows smoothly towards $x=0$, its
velocity increasing in time. While $t$ approaches $t_{\rm cr}$,
more and more fluid keeps coming in,  causing the density gradient
$(\partial/\partial x)\rho(x=0, t_{\rm cr})$ to blow up. This
shows as a cusp in the plot of $\rho(x, t_{\rm cr})$. After that,
the fluid starts accumulating at $x=0$, and the velocity $v(x, t)$
becomes smooth again. 

\ifig\shocksfigure{
The droplet density $\rho(x, t)$ at various stages of evolution:
$t<t_{\rm cr}$ (dotted line), $t=t_{\rm cr}$ (solid line) and 
$t>t_{\rm cr}$ (dashed line).
The plot of $\rho(x, t_{\rm cr})$ exhibits a ``shock'' at $x=0$.}
{\epsfysize2.5in\epsfbox{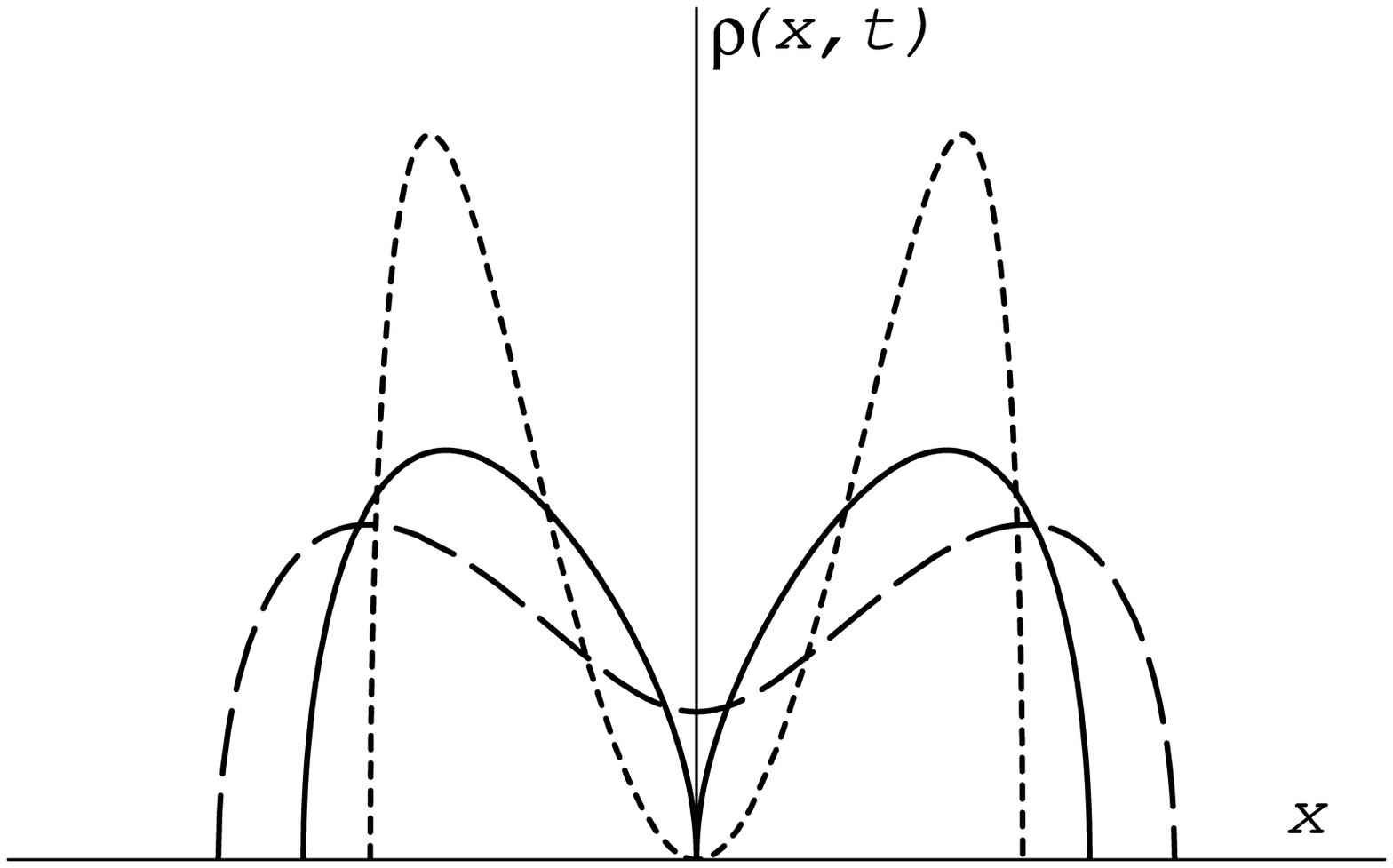}}

Nothing similar ever occurs in the $\epsilon<1$ phase. There at 
$g=g_{\rm cr}(\epsilon)$ the droplet density behaves as $\rho(x, t)\propto
\alpha(t) |x|+ {\cal O}(x^2)$ with $\alpha(t)$ being an infinitely smooth 
function of time. In that phase $\rho(x=0, t)\equiv 0$ for all $t\in[0,1]$.

One could say that the quantity $\zeta=\rho_{1/2}(0)$ is an ``order
parameter'' for our theory. Indeed, $\zeta\equiv 0$ identically at any 
$\epsilon<1$. On the other hand, when $\epsilon>1$ the values of $\zeta$ 
are manifestly positive.
Although such an order parameter does not describe the spontaneous breaking 
of any symmetry, it does parametrize a change in the group representation 
structure of our model. This shall be demonstrated in section~4. 
But before that let us use the intuition derived from the large $\epsilon$
limit to
construct an exact solution of the infinite matrix chain at $\epsilon>1$.

\newsec{Exact Solutions of the Matrix Chain for Lattice Spacings Greater
than One}

In this section we shall obtain a class of potentials $U(x)$ for which the
chain model can be solved exactly. We shall see that
the critical index $\delta$ corresponding to these  potentials 
equals $\delta(\epsilon)=2$ at any $\epsilon$ 
between one and infinity.

The solutions presented below are not as explicit as exact solutions
are usually expected to be. Neither the interaction potentials nor the 
eigenvalue densities corresponding to these solutions can be written
in terms of elementary or special functions. Instead, they are determined
by complicated transcendental equations, although the critical indices 
for the ``transcendental'' solutions are still calculable explicitly.

In principle, to produce such exact solutions is not hard. Choose 
an arbitrary, but properly normalized $\rho_{1/2}(x)$.  Imagine a droplet
which at $t=1/2$ is at rest and has this density. Evolve this droplet 
backwards in time up to $t=0$ and read off the fluid velocity $v(x)=v(x, t=0)$
together with the density $\rho(x)=\rho(x, t=0)$. The result is an exact 
solution---given by $\rho(x)$\ ---to the matrix chain model where the potential
$U(x)$ can be found from \initvelocity.

The most difficult part of this program is to adjust $\rho_{1/2}(x)$ so that
the potential $U(x)$ comes out physically reasonable. Let us require that
$U(x)$ be a smooth and bounded from below double-well potential with
a hump at $x=0$,
\eqn\humppot{
U(x)=-{\mu^2\over 2} x^2 +{\cal O}(x^4).}
Here, as before, $\mu^2=2(2\epsilon^2-1)$.

The potentials in this class are qualitatively similar to the pure quartic 
potential \pott. Therefore, due to the phenomenon of universality, we can 
expect that the critical exponents of the chain model with these potentials
are strictly equal to the critical exponents of the pure quartic model.
This hypothesis shall be confirmed in section 5 where we investigate
the pure quartic case using a systematic strong coupling expansion.

A carelessly chosen $\rho_{1/2}$ would most likely yield the potential
with $\mu^2<2$, which corresponds to $\epsilon<1$. 
To generate $U(x)$ with $\mu^2>2$ the analytic structure 
of $\rho_{1/2}$ must be very special.
This structure can be inferred from the results for the $\epsilon\gg 1$ limit
obtained in section 2. That is, let us choose $\rho_{1/2}(x)$ so that\hfill
\break
(a) as an analytic function of $x$, it has at least two branches (not counting 
the branches different only by a minus sign);\hfill\break
(b) one of these branches is an even function of $x$, real valued for 
real $x$ (this shall be the ``physical'' branch representing the actual
droplet density at $t=1/2$);\hfill\break
(c) the ``physical'' branch is normalized, $\int\rho_{1/2}(x)\, dx=1$;
\hfill\break
(d) the second branch of $\rho_{1/2}$ has the Taylor expansion given by
\taylor.


Although these requirements are quite restrictive, there are many functions
obeying all of the conditions (a)--(d). As an example, consider the 
function $r(x)=[\pi\rho_{1/2}(x)]^2$ specified by the following
parametric representation 
\eqn\parametricfunction{
\left\{\eqalign{
r(i\xi)&=R^2(\varphi)\cosh(\kappa \varphi)\cr
\xi&=R(\varphi)\sinh(\kappa \varphi)\cr}\right.}
where $\varphi\in[0, \pi/2]$ and $R(\varphi)$ is an even analytic function
of $\varphi$ such that $R(\pi/2)=0$. For concreteness, let us choose
\eqn\Rvarphi{
R(\varphi)=\sum_{n=1}^{n_{\rm max}} R_n \cos^n\!\varphi}
with arbitrary real coefficients $R_n$.

It is easy to check that the plot of $r(i\xi)$ has two branches as in fig.~3
and that the ``physical'' branch of $r$ is even. Furthermore, the Taylor
expansion of the ``hidden'' branch has the form which agrees precisely 
with \taylor,
\eqn\taylorexample{
r(i\xi)=\xi^2 {\cosh(\pi\kappa/2)\over \sinh^2(\pi\kappa/2)}
\Biggl\{1+ \xi {\kappa\over R_1}
{\cosh(\pi\kappa/2)\over \sinh^2(\pi\kappa/2)}\biggl[2-\tanh^2\biggl({\pi\kappa
\over 2}\biggr)\biggr]+ {\cal O}(\xi^2) \Biggr\}}
so that condition (d) is also satisfied.
The values of the physical branch at real $x$ can be found from 
\parametricfunction\ using the substitution $\varphi=-i\eta$,
\eqn\physvalues{\left\{\eqalign{
r(x)=&\cos (\kappa\eta) \Biggl[\sum_{n=1}^{n_{\rm max}} R_n 
\cosh^n\!\eta\Biggr]^2\cr
x=&\sin(\kappa\eta)\sum_{n=1}^{n_{\rm max}} R_n 
\cosh^n\!\eta.\cr}\right.}
As we see, the function $r(x)$ vanishes at $\eta=\pm\pi/2\kappa$.
Therefore, the support of $\pi\rho_{1/2}(x)=\sqrt{r(x)}$ is restricted to the
finite interval $[-x_{\rm max}, x_{\rm max}]$ with $x_{\rm max}=x(\eta=
\pm\pi/2\kappa)$, and therefore the
normalization integral $\int\rho_{1/2}(x)\,dx$ is
finite. The value of this integral can be adjusted to equal one by
rescaling all the coefficients $R_n\to \lambda R_n$
with an appropriately chosen $\lambda$.

The potential $U(x)$ which corresponds to \parametricfunction\
must be a nonsingular infinitely differentiable function of $x$. This
condition should be enforced at least for those $x$ where $\rho(x)\ne 0$\
---otherwise additional spurious singularities in $\rho(x)$ could be produced
and the universality argument (the paragraph after \humppot) would become 
inapplicable.

Technically, $U$ and $\rho$ are determined by the $t=0$ values of the Hopf 
function $f(x, t)=v(x, t)+i\pi\rho(x, t)$ which, in turn, is constrained by 
equation \implicithalfback. Consequently, $f$ solves a system of 
transcendental equations obtained from \physvalues\
by the substitution
\eqn\substitution{
\left\{\eqalign{
x&\to x+\tau f(x, t)\cr
r&\to -f^2(x, t).\cr}\right.}
Generally, this equation has multiple roots. As $\tau=1/2-t$ increases
from $0$ to $1/2$, some of these roots might coalesce producing 
unwanted singularities in $U$ and then also in $\rho$. Therefore, the
parameters $R_n$ in \Rvarphi\ should be chosen so that this does not
happen for any $\tau<1/2$.

The numerical analysis of such transcendental equations indicates
(although we do not have a rigorous proof) that we do have some choice of 
parameters
for any $\epsilon>1$.  This immediately allows us to determine 
the critical index $\delta$ in the $\epsilon>1$ phase of the matrix chain.
Indeed, using the Taylor expansion \taylorexample\ we can read off
the coefficients
\eqn\coefficients{\left\{
\eqalign{
a_1(\epsilon)&=\sqrt{{\cosh(\pi\kappa/2)\over \sinh^2(\pi\kappa/2)}}\cr
a_2(\epsilon)&={\kappa\over 2R_1}\biggl[{\cosh(\pi\kappa/2)\over 
\sinh^2(\pi\kappa/2)}\biggr]^{3/2}\biggl[2-\tanh^2\biggl({\pi\kappa
\over 2}\biggr)\biggr].\cr}\right.}
From \aone\ we now deduce
that the parameter $\kappa$ is a function of $\epsilon$,
\eqn\kappaepsilon{
{\cosh(\pi\kappa/2)\over \sinh^2(\pi\kappa/2)}=\biggl({2\epsilon^2\over 
\epsilon^2-1}\biggr)^2}
and that for any lattice spacing $a_2(\epsilon)>0$.
Consequently, the eigenvalue density in the chain model behaves as 
$\pi\rho(x)\equiv {\rm Im}\,f(x, t=0)\propto f_2 x^2$, where 
$f_2>0$, yielding the eigenvalue index in the $\epsilon>1$ phase $\delta
(\epsilon)=2$.

By replacing $\cosh(\kappa\varphi)$ and $\sinh(\kappa\varphi)$ in
\parametricfunction\ with more complicated functions it is possible to
obtain models where at certain discrete values of the lattice spacing
$\epsilon$ the coefficient $a_2(\epsilon)$ vanishes. At these $\epsilon$
such models exhibit higher order multicritical behavior. Of course,
the existence of multicritical points in the infinite matrix chain should not
come as a surprise. Indeed, exactly the same spectrum of critical indices 
occurs in one-matrix models which describe the $\epsilon\gg 1$ limit
of the chain theory.

In the case of the upside-down quartic potential $V(M)=m^2 M^2/2+{\tilde g}
M^4/4$ with $m^2>0$ and ${\tilde g}<0$ the $\epsilon\gg1$ limit 
of the matrix chain becomes a $c=0$ theory characterized by $\delta=3/2$.
Using the arguments of this and the previous sections
it is possible to show that there exist regular potentials leading to
$\delta(\epsilon)=3/2$ at any $\epsilon>1$. Again, the midway density
$\rho_{1/2}(x)$ develops a second branch when the lattice spacing $\epsilon$
gets bigger than one, while only one branch enters the picture for
$\epsilon<1$. However, to see the second branch one should continue 
$\rho_{1/2}(x)$ not to imaginary $x=i\xi$ but rather to real
positive  very large $x\gg1/m$. Since for real $x$ the density
$\rho_{1/2}(x)$ must always be real, the corresponding ``hidden''
branch can only have
half-integer singularities $\rho_{1/2}(x)\propto (x-x_0)^{\delta}$
with $\delta=n+1/2$, the lowest nontrivial value of $\delta$
being $\delta=3/2$. 

\newsec{The Character Expansion and the Order Parameter}

In this section we shall develop the character expansion for the chain model.
We shall find that the $SU(N)$ representation dominating the chain
partition function at large $N$ develops a ``gap'' at the Kosterlitz--Thouless
phase transition point, and that 
the gap size is a simple function of the order 
parameter $\zeta=\rho_{1/2}(0)$.

There are both conceptual and technical reasons why the character 
expansion is useful in the matrix chain problem. Conceptually, it shows
that the Kosterlitz--Thouless phase transition and the Douglas--Kazakov-type
transitions in two-dimensional QCD and the dually weighted graph models
all occur through essentially the same mechanism. On a technical level, 
there is some evidence, based on the singularity structure of the matrix chain
with $\epsilon<1$, that the exact solution of the chain model with a
pure quartic potential may belong to the class of elliptic 
functions \REFfirstpaper.
If this is true, the character expansion could be just the right 
computational tool to derive such a solution.

To construct the character expansion for the infinite matrix chain, let us 
go back to the partition function \chaintwo\ and diagonalize each matrix 
$\calm_n=U_n\Lambda_n U_n^{\dagger}$ with diagonal 
$\Lambda_n={\rm diag}(\lambda_{1,n}\dots \lambda_{N,n})$.
Then, 
express the Hermitian matrix measure $d\calm_n$ in terms of $U$'s and
lambdas,
$$d\calm_n=\Delta^2(\lambda_{n}) \thinspace d\lambda_{1,n} \dots 
d\lambda_{N,n} \thinspace dU_n$$
where $\Delta(\lambda_n)$ is the Van der Monde determinant of eigenvalues
$\lambda_{i,n}$ and $dU_n$ refers to the Haar measure on the unitary group
$SU(N)$. Furthermore, it is useful to introduce the matrices 
$V_n=U_{n+1}^{\dagger}U_n$.

Given this notation the partition function of the infinite matrix chain
can be written in the form 
\eqn\chaindiag{
{\cal Z}=\!\!\int\!\prod\limits_{n\in \IZ}^{}\!\!\Delta^2(\lambda_n)
 \, d\lambda_{1,n} \dots 
d\lambda_{N,n} dU_n \, \exp\biggl\{N \, {\rm tr}\!\!
\sum_{n=-\infty}^{\infty}
\Bigl[ V_n \Lambda_n V_n^{\dagger} \Lambda_{n+1} - U(\Lambda_n)\Bigr]\biggr\}.}
The matrices $V_n$ represent the angular degrees of freedom so typical
of all multimatrix models.
Fortunately, since the one-dimensional lattice does not have closed loops,
all $V_n$ are mutually independent and we can easily integrate them out.
To do this one simply changes variables from $\{U_n\}$ to $\{V_n\}$ 
with the result
\eqn\chaindiagtwo{
{\cal Z}=\!\!\int\!\prod\limits_{n\in\IZ}^{}\Delta^2(\lambda_n)
\, d\lambda_{1,n} \dots 
d\lambda_{N,n} dV_n \, \exp\biggl\{N \, {\rm tr}\!\!\sum_{n=-\infty}^{\infty}
\Bigl[ V_n \Lambda_n V_n^{\dagger} \Lambda_{n+1} - U(\Lambda_n)\Bigr]\biggr\}.}
Note that no Jacobian arises when we pass from $U$ to $V$.

In the large $N$ limit the partition function \chaindiagtwo\ can be 
simplified even further. Indeed, imagine that all integrals over $V_n$
have been done. Then for $N\to\infty$ 
the remaining integrals over the eigenvalues are dominated by 
an $n$-independent
saddle point $\lambda_{i, n}\equiv\lambda_i$. Remarkably, the same saddle 
point values of $\lambda_i$ occur in a simpler integral
\eqn\eguchi{
{\cal Z}_{\rm EK}=\int dV\, \Delta^2(\lambda) d\lambda_1\dots d\lambda_N
\, \exp\Bigl\{N\, {\rm tr}\big[V \Lambda V^{\dagger} \Lambda - U(\Lambda)
\big]\Bigl\}}
where 
$\Lambda={\rm diag}(\lambda_1, \dots, \lambda_N)$. To prove this
assume  
$$\int dV \, \exp\big[N\, {\rm tr}(V \Lambda V^{\dagger} 
\Lambda^{\prime})\big]=\exp\big[N^2F(\Lambda, \Lambda^{\prime})\big]$$ 
with a certain
$F(\Lambda, \Lambda^{\prime})$, and write down the saddle point equations
for both integrals. These equations happen to be the same, and so are their 
solutions.
In fact, the equivalence between ${\cal Z}$ and ${\cal Z}_{\rm EK}$ is just
a very special case of the Eguchi--Kawai reduction theorem \REFEKreduction.

There are several ways to integrate out the matrix $V$ in \eguchi.
One of them---based on the large $N$ asymptotics of the famous Itzykson--Zuber 
integral \REFhopf\REFitzyksonzuber---leads eventually 
to the hydrodynamic picture \eqnstate--\returninitial.
Another useful method is the character expansion \REFwynter.
Instead of trying to do the integral over $V$ explicitly, we expand the
integrand in a series of $SU(N)$ characters and then use the formula
\eqn\rusakov{
\int dV \, \chi_R(VAV^{\dagger}B)= {1\over d_R}\chi_R(A)\chi_R(B).}
Here  $\chi_R(U)$ stands for the character of a unitary matrix $U$ in the
representation $R$; $d_R$ is the dimension of that representation, while
$A$ and $B$ are arbitrary matrices.
The final output of a character expansion is a sum over representations of 
$SU(N)$ which can usually be done by the saddle point method \REFwynter.

When dealing with large $N$ limits, it is often more convenient to
expand in the characters of $U(N)$ rather than $SU(N)$. The $U(N)$
representations are labelled by $N$ integers
$n_1\ge\dots\ge n_N$, the character of a matrix ${\cal M}$ in any such
representation being  
given by the Weyl formula
\eqn\weyl{
\chi_R({\cal M})={\det\bigl|\!\bigl|\lambda_p^{l_q}\bigr|\!\bigr|
\over \Delta(\lambda)}.}
As usual, the numbers $\lambda_1, \dots, \lambda_N$ denote the eigenvalues of
the matrix ${\cal M}$, whereas the (strictly ordered) integers $l_i$
equal $l_i=n_i+N-i$.

To compute the integral \eguchi\ let us expand the term containing
$W=V\Lambda V^{\dagger}\Lambda$ in a character series \REFitzyksonzuber\
\eqn\expans{
{\rm e}^{N \, {\rm tr}\, W}=\sum_R c_R \, \chi_R(W).}
The coefficients of this expansion are easily found from the
orthogonality of characters,
\eqn\computation{
\eqalign{
c_R=&\int dV {\rm e}^{N \, {\rm tr}\, V} \chi^*_R(V)\cr
=& {\Delta(l)\over l_1!\dots l_N!} \, N!\,  
N^{(l_1+\dots +\l_N)-N(N-1)/2}.\cr}}
Now we can apply \rusakov\ to do the unitary group integration in 
\eguchi. Throwing out a constant overall factor and using 
the formula for the dimension of a 
representation $R=\{l_1, \dots, l_N\}$
\eqn\dimension{
d_R=\prod\limits_{i<j} {l_i-l_j\over j-i}}
this finally yields
\eqn\chexpan{
{\cal Z}_{\rm EK}=\sum_{l_1, \dots, l_N=0}^{\infty}{N^{l_1+\dots +\l_N}
\over l_1!\dots l_N!} \int  d\lambda_1\dots d\lambda_N\,\Delta^2(\lambda)
\, \chi_R^2(\Lambda) \,  {\rm e}^{-N \, {\rm tr}\,U(\Lambda)}.}
At large $N$ both the sum over $l_1,\dots,\l_N$ and the integral over the 
eigenvalues in \chexpan\ are dominated by saddle points. Formally, these saddle
points are defined as the values of $l_1,\dots,\l_N$ and $\lambda_1, \dots
\lambda_N$ that maximize the integrand of \chexpan.
In fact, the saddle point for $\{\lambda_i\}$ is already known to us. It is
given by the same eigenvalues that dominate the original 
matrix chain integrand \chaintwo\ at $N\to\infty$ and which are described by 
the density $\rho(\lambda)$ in \returninitial.

At the saddle point in the representation space the values of $l_i$
are all of order $N$. Therefore, it shall be convenient to introduce the
quantities $h_i=l_i/N$ which, in the large $N$ limit, condense to form a
smooth distribution $\rho_l(h)$. Since all of the $l$'s are different from
each other, this density can never be greater than one.

To maximize the integrand of \chexpan\ with respect to $l_1,\dots,\l_N$ we 
first use the Stirling formula $N^{l_i}/l_i!\approx\exp[N h_i(1-\log h_i)]$.
Also, the large $N$ asymptotics of the character $\chi_R(\Lambda)$
is given by $\chi_R(\Lambda)\propto \exp\{N^2\Xi[\rho_l(h), \rho(\lambda)]\}$
where the functional $\Xi[\rho_l(h), \rho(\lambda)]$ has a finite
large $N$ limit. In view of this notation, the saddle point equation
for the $l$'s can be written in the form
\eqn\saddreps{
2{\partial\over \partial h} {\delta\over \delta \rho_l(h)}
\Xi[\rho_l(h), \rho(\lambda)]=\log h.}
The equations needed to compute $\Xi$ are provided by the method of \REFhopf.
However, we should be particularly careful as some of the $\lambda_i$
(in fact, half of them for an even $U(\lambda)$) have a negative sign.
Then the sign of an expression like $\lambda_p^{l_q}=\lambda_p^{Nh_q}$ 
oscillates when $N$ increases and its large $N$ behavior is not well
defined. To circumvent the difficulty, imagine that $\lambda_1, \dots,
\lambda_{N/2}$ are all positive and that the negative ones are just
$-\lambda_1, \dots,-\lambda_{N/2}$. Furthermore, assume that among $l_i$
half are odd (call them $l_i^{\rm o}$) and half are even (denoted 
$l_i^{\rm e}$.) Certainly, if the set of all (odd and even) $l$'s
converges to form the distribution $\rho_l(h)$, the same distribution
would also describe the densities of both $\{l_i^{\rm o}\}$ and
$\{l_i^{\rm e}\}$ taken separately from each other.

For such symmetric distributions of $\{\lambda_i\}$ and $\{l_i\}$
the Weyl determinant factorizes into a product of
two $(N/2)\times(N/2)$ determinants \REFwynter\
\eqn\deters{
\det\big|\!\big|\lambda_p^{l_q}\big|\!\big|=
(-2)^{N/2} \det\big|\!\big|\lambda_{+p}^{l_q^{\rm e}}\big|\!\big|
\det\big|\!\big|\lambda_{+p}^{l_q^{\rm o}}\big|\!\big|}
where $\lambda_{+p}\, (p=1,\dots, N/2)$ are the $N/2$ positive eigenvalues
$\lambda_1, \dots,\lambda_{N/2}$.
Both of these smaller determinants exhibit identically the same leading
large $N$ behavior  $\exp\{N_*^2 {\cal D}[\rho_l(h), \rho(\lambda)]\}$
with $N_*=N/2$ being the new determinant size.

To find ${\cal D}$ we first represent $\lambda_{+p}$ in the manifestly 
positive form $\lambda_{+p}=\exp(\theta_p/2)$ so that 
$\lambda_{+p}^{l_q^{\rm e}}=\exp(N_*\theta_p h_q^{\rm e})$.
The density of thetas $\sigma(\theta)$ is, of course, related to
$\rho(\lambda_+)$. Since $N_*\sigma(\theta)\, d\theta=N\rho(\lambda_+)\, 
d\lambda_+$, one gets
\eqn\sigmarel{
\sigma(\theta)={\rm e}^{\theta/2}\rho({\rm e}^{\theta/2}).}
Now we can write down the equations which fix ${\cal D}$. Let us introduce
the two functions
\eqn\gplusminus{
\left\{\eqalign{
G_+(\theta)&= {\partial\over \partial \theta}{\delta{\cal D}\over\delta
\sigma(\theta)}+i\pi\sigma(\theta)\cr
G_-(h)&={\partial\over \partial h}{\delta{\cal D}\over\delta
\rho_l(h)}-i\pi\rho_l(h).\cr}\right.}
Then the equation on ${\cal D}$ is simply \REFhopf\REFgrosscylinder\
\eqn\myusual{
G_+\big[G_-(h)\big]=h.}
This complex-valued functional equation is equivalent to two real-valued
constraints which determine the two unknown functions 
$(\partial/\partial \theta)
{\delta{\cal D}/\delta
\sigma(\theta)}$ and 
$({\partial/\partial h}){\delta{\cal D}/\delta
\rho_l(h)}$.

On the other hand, the functional derivatives of ${\cal D}$ can be extracted
from the saddle point equations. In this way, equation \saddreps\ yields
\eqn\fderl{
{\partial\over \partial h}{\delta{\cal D}\over\delta
\rho_l(h)}=\log h.}
We can also replace $\lambda_i\to\pm\exp(\theta_i/2)$
in \chexpan, maximize it with respect to $\theta_i$ 
and get the saddle point equation for
the thetas. Remembering that $N=2N_*$, we derive
\eqn\fdert{
{\partial\over \partial \theta}{\delta{\cal D}\over\delta
\sigma(\theta)}={1\over 2}{\rm e}^{\theta/2} U^{\prime}({\rm e}^{\theta/2}).}
Finally, let us combine \gplusminus, \fderl\ and \fdert\ together with
equation \myusual. The result can be conveniently expressed in terms 
of an auxiliary function $H(z)=U^{\prime}(z)/2 +i\pi\rho(z)$
and reads 
\eqn\eich{
H\biggl[\sqrt{h}\,\exp\biggl(-{i\pi \rho_l(h)\over 2}\biggr)\biggr]
=\sqrt{h}\, \exp\biggl({i\pi \rho_l(h)\over 2}\biggr).}
Very remarkably, this equation leads to a very simple 
formula connecting $\rho_l$ to $\rho_{1/2}$. Indeed,
from \initvelocity\ and \implicitforward\ we have the following
relation between $\rho_{1/2}$ and $H$, 
\eqn\connect{
i\pi\rho_{1/2}\Bigl[\half\big(H(x)+x\big)\Bigr]=H(x)-x.}
Now let us substitute $x=\sqrt{h}\exp[-i\pi\rho_l(h)/2]$
and use \eich. This produces, at last, the formula
\eqn\dua{
\pi \rho_{1/2}\biggl[\sqrt{h}\,\cos\biggl({\pi \rho_l(h)\over 2}\biggr)\biggr]
=2\sqrt{h}\, \sin\biggl({\pi \rho_l(h)\over 2}\biggr)}
which is our final result.

\ifig\tableaudensitiesfigure{The Young tableau densities $\rho_l(h)$
for $\epsilon<1$ (left) and for $\epsilon>1$ (right). The presence of a gap in
the Young tableau for $\epsilon>1$
shows as a plateau on the graph of $\rho_l(h)$.} 
{\epsfxsize2.0in\epsfbox{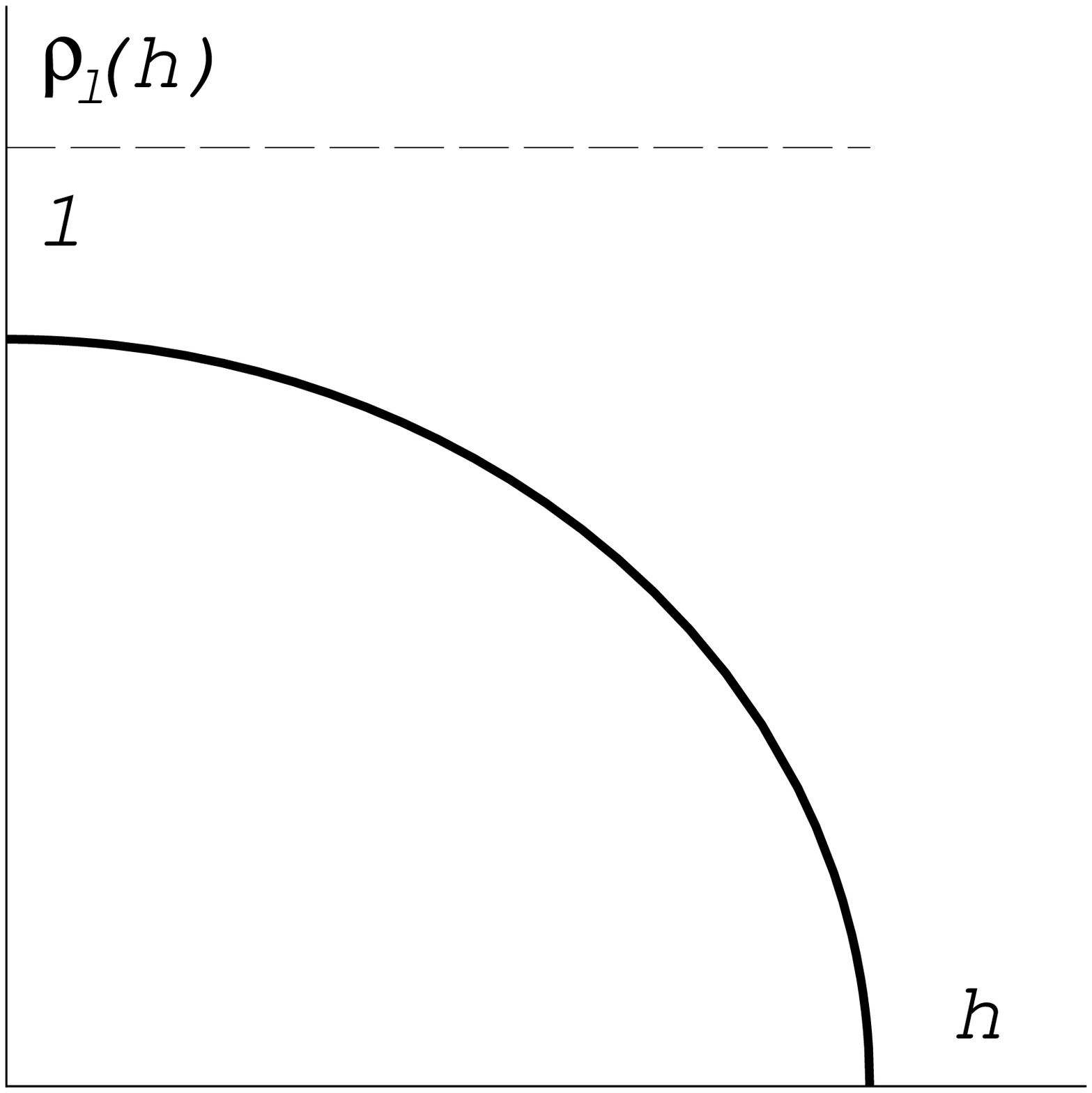}
\hskip0.5in\epsfxsize2.0in\epsfbox{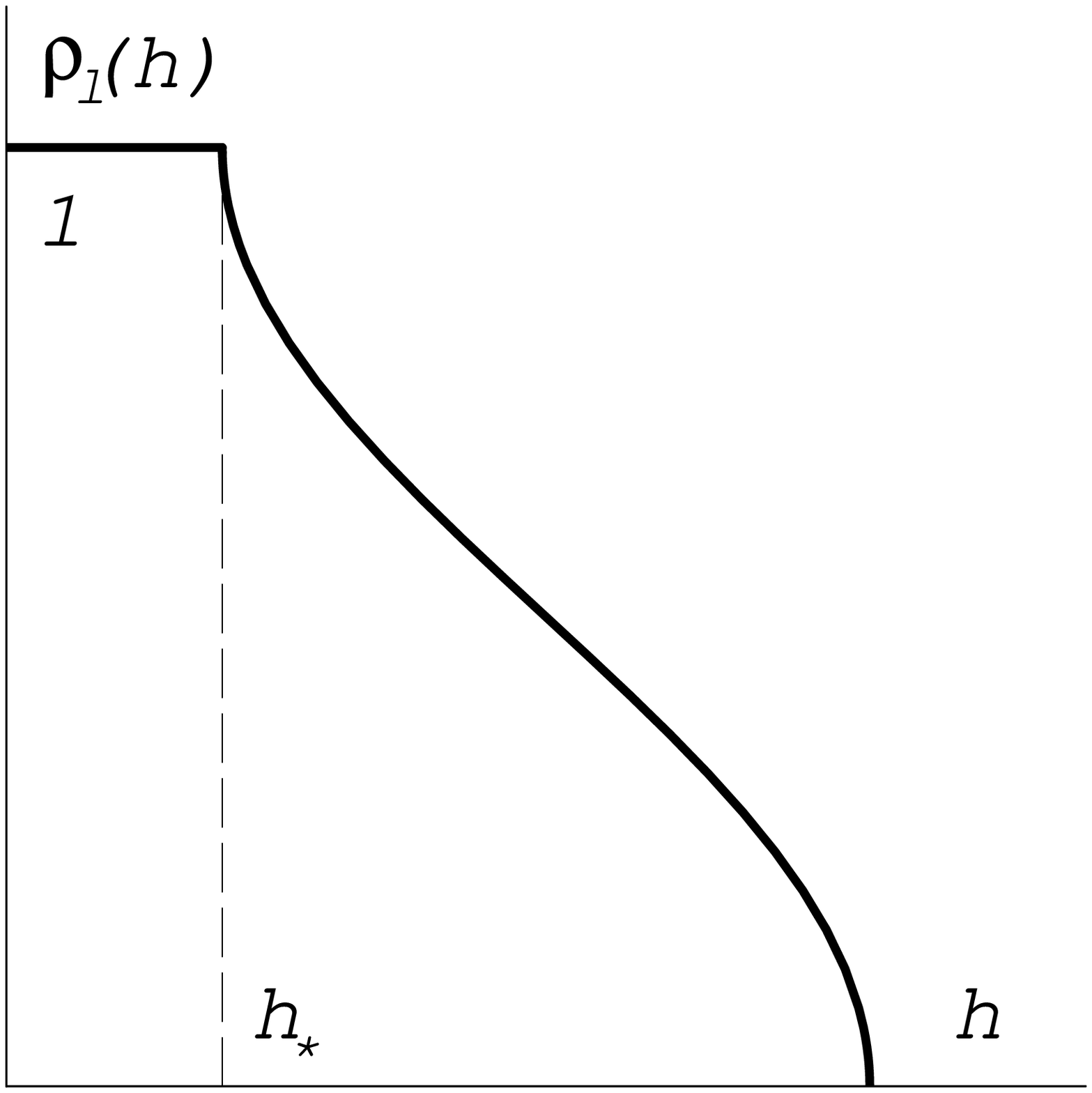}}

Equation \dua\ is essentially kinematical. Indeed, it does not involve the 
matrix model interaction potential in any way\foot{Equation \dua\ can also
be derived in an ``explicitly kinematic'' fashion 
following the ideas of Douglas \REFdouglas.}.
However, it does indicate that the descriptions of the infinite chain model
based on the hydrodynamic picture and on the character expansion are
completely equivalent.

To be rigorous, equation \dua\ holds only for those $h$ where $\rho_l(h)$
is strictly less than one. Indeed, if the Young tableau density reaches
its upper bound $\rho_l(h)=1$ the saddle point equation \fderl\ ceases
to apply and the whole derivation becomes invalid. Practically,
this allows us to determine whether there is, in fact, an interval of
$h$ where $\rho_l(h)=1$.  If yes, then the size of this interval $h_*$
equals the ``gap width'' in the corresponding Young diagram as shown in
fig.~1.

%

Since $h_*$ is just on the border of the region where \dua\ applies, we can 
use that formula with $h=h_*$ and $\rho_l(h_*)=1$. This yields
$\zeta=\pi\rho_{1/2}(0)=2\sqrt{h_*}$ thereby proving equation \zetaeqn.
In other words, the gap size, which appears to be a natural ``order parameter''
for the phase transition in the character expansion language is related 
to $\zeta$ ---an order parameter in the hydrodynamic picture.

Therefore, in the critical regime $h_*\equiv 0$  for $\epsilon<1$ while
$h_*=h_*(\epsilon)>0$ for any $\epsilon>1$. This picture of the 
Kosterlitz--Thouless phase transition is completely identical to the 
mechanism of the Douglas--Kazakov transition in large $N$ two-dimensional
QCD \REFdouglaskazakov\REFgrossinstantons. 
Indeed, both in the matrix chain model and in QCD$_2$ the 
density $\rho_l(h)$ develops a plateau for large lattice spacings 
(respectively, strong coupling) while $\rho_l(h)<1$ everywhere for small
lattice spacings (or weak couplings.)
And, to complete the correspondence, both transitions are induced by the 
topologically nontrivial states---vortices and instantons---that dominate
in the strongly coupled phases of these two theories.

\newsec{The Chain Model with a Pure Quartic Potential}

In this section we shall construct and explore a systematic strong 
coupling expansion for the infinite matrix chain with a quartic
interaction potential.
The expansion we shall consider is in powers of $1/\epsilon^2$ or, 
equivalently, $1/\mu^2$.

For large $\mu^2$ the successive terms in the character expansion
\chexpan\ become smaller and smaller. Indeed, any character $\chi_R(\Lambda)$
is a homogeneous polynomial in the eigenvalues $\lambda_i$. The degree of
homogeneity for such a polynomial equals the total number of boxes in the 
corresponding Young tableau, $n_R=n_1+\dots +n_N=l_1+\dots+l_N-N(N-1)/2$.
Consequently, if for large $\mu$ we approximate the eigenvalue density
$\rho(\lambda)$ by the one-matrix model result \onematstrong, the 
typical width of such distribution will be $\lambda_{\rm typ}\sim 1/\mu$
and the value of the character $\chi_R(\Lambda)$ will be suppressed by 
a factor $1/\mu^{n_R}$. In other words, to expand in $1/\mu$ up to
the order $1/\mu^{2n_R}$ we should choose only those representations in 
\chexpan\ which have not more than $n_R$ boxes. The characters in
such representations shall be polynomials in ${\rm tr}\Lambda, 
{\rm tr}\Lambda^2, \dots, {\rm tr}\Lambda^{n_R}$ and would effectively generate
corrections to the one-matrix potential $U(\lambda)$. That is to say, 
they will perturb the right hand side of the Riemann--Hilbert equation
\riemannhilbert\ by the terms of order at most $\lambda^{n_R-1}$.

A one-matrix model with such modified potential has, in general, the 
solution\foot{Since we are interested in the critical regime, the 
coupling constant $g$ is always adjusted  to equal $g_{\rm cr}(\epsilon)$.
For the uncorrected one-matrix model this would imply $g=\mu^4/4$.}
\eqn\rhopolyn{
\pi\rho(x)=Q(x)\sqrt{1-{\mu_{\rm r}^2 x^2\over 8}}}
where $Q(x)$ is a polynomial of degree $n_R-2$ and $\mu_{\rm r}$ ---the 
``renormalized'' distribution width \REFbipz. 
Obviously, if the original potential 
$U(\lambda)$ was even, so shall be $Q(x)$. The coefficients of $Q$
could, in principle, be determined by computing the characters, evaluating
the corrections to $U(\lambda)$ and solving the resulting one-matrix model.

Fortunately, there is a more economical way to organize this computation.
Let us use the fact \REFhopf\REFfirstpaper\
(following essentially from \implicit) that for any 
infinite chain model the functions 
\eqn\gpmchain{
{\cal G}_{\pm}(z)={1\over 2}U^{\prime}(z)\pm i \pi\rho(z)}
(with $\rho(z)$ being the exact eigenvalue density) obey the functional
equation
\eqn\func{
{\cal G}_+\big[{\cal G}_-(z)\big]=z.}
Very remarkably, this equation allows one to compute $Q(z)$
much easier and quicker.

The left hand side of equation \func\ contains the
eigenvalue density at a complex point, $\rho[{\cal G}_-(z)]$. Consequently,
it is important to choose the correct analytic branch of $\rho$ in both
${\cal G}_+$ and ${\cal G}_-$. Since it is more convenient to work with real
variables, we shall enforce equation \func\ in the region of large
positive $z>2\sqrt{2}/\mu_{\rm r}$, where $i\pi\rho(z)$ becomes real.
The correct sign choice for $\rho(z)$ in that region yields
\eqn\gcontinued{
{\cal G}_+(z)={\cal G}_-(z)\equiv {\cal G}(z)=-{\mu^2\over 2} z+{g\over 2}z^3
-Q(z){\mu_{\rm r} z\over 2\sqrt{2}}\sqrt{1-{8\over \mu_{\rm r}^2 z^2}}.}
The reason is, $\rho(z)$ has a square root branch cut between $x_{\rm max}=
2\sqrt{2}/\mu_{\rm r}$ and $-x_{\rm max}=-2\sqrt{2}/\mu_{\rm r}$. If 
a real $z$ were positioned on the upper edge of this cut the value of
${\cal G}_-(z)$ would have a negative imaginary part. 
In evaluating $\rho[{\cal G}_-(z)]$ 
this would take us below the lower edge. But the values of $\rho(z)$ 
on the lower edge are opposite to the values of $\rho$ on the upper
edge. When $z$ is continued to $z>2\sqrt{2}/\mu_{\rm r}$ 
the branch of $\rho(z)$
entering ${\cal G}_-$ should be continued from above the cut while 
$\rho(z)$ in ${\cal G}_+$ should be continued from below the cut. 
This compensates
for the sign difference in front of $\rho(z)$ in \gpmchain\ and leads to the
result \gcontinued.

At infinitely large $\mu$, which corresponds to the one-matrix model case,
we get
\eqn\gonemat{
{\cal G}_+(z)={\cal G}_-(z)={1\over z}+{\cal O}\biggl({1\over \mu^2}\biggr)}
so that equation \func\ is trivially satisfied.
The $1/\mu^2$ corrections can now be derived by perturbing this solution.
Since we are interested in the critical regime where 
$\rho(x)\propto |x|^{\delta}$ the polynomial $Q(z)$ must vanish at $z=0$,
and we can seek a solution of the form
\eqn\figexp{
{\cal G}(z) = - {\mu^2 \over 2}z + {g \over 2} z^3 -
 z^3 P(z) {\mu_{\rm r}^4 \over 8} \sqrt{1- {8 \over \mu_{\rm r}^2 z^2}}.}
The coupling $g=g_{\rm cr}(\mu)$ and the coefficients of $P(z)$ should 
be computed
from equation \func\ together with the normalization requirement $\int\rho(x)\,
dx=1$. 

Technically, it is easier to expand in powers of $1/\mu_{\rm r}^2$ rather than 
$1/\mu^2$. This is perfectly legitimate given that, to the leading order
in our large $\mu$ expansion, $\mu=\mu_{\rm r}$.
In other words, we shall seek corrections to the one-matrix model relations
$\mu=\mu_{\rm r}$ and $g_{\rm cr}=\mu_{\rm r}^4/4$,
\eqn\figmu{\left\{
\eqalign{
\mu^2 & = \mu_{\rm r}^2\biggl(1+ \sum_{n=1}^{M+1} {b_n \over 
\mu_{\rm r}^{2n}}\biggl) \cr
g_{\rm cr} & = {\mu_{\rm r}^4\over 4} \biggl(1+ \sum_{n=1}^{M+2} 
{g_n \over \mu_{\rm r}^{2n}}\biggl)\cr}\right.}
where we cut the expansion at order $1/\mu_{\rm r}^{2M}$. For the
polynomial $P(z)$ we assume 
\eqn\fiP{
P(z) = 1+ {1\over \mu_{\rm r}^4} \sum_{i=1}^{M} {e_i \over \mu_{\rm r}^{2i}} +
{1\over \mu_{\rm r}^8} \sum_{s=1}^{M-2} {z^{2s}\over \mu_{\rm r}^{2s}}
\sum_{i=1}^{M-1-s} {a_{s,i}\over \mu_{\rm r}^{2i-2}}.}
Note that, to ${\cal O}(1/\mu^{2M})$, the degree of $P(z)$
equals $2(M-2)$, fully consistent with \rhopolyn.

To determine the coefficients $b_n, g_n, e_i$ and $a_{s,i}$ we impose 
${\cal G}[{\cal G}(z)]=z+{\cal O}(\mu^{-2M})$ and require 
the normalization of the density to be $1+
{\cal O}(\mu^{-2M})$. Practically, the normalization requirement  
amounts to asking that the coefficient of $1/z$ in the large $z$ 
expansion of \figexp\ equal one.
The resulting 
system of equations can be solved recursively using the following
two stage
procedure. First, one expresses $b_{M+1}, g_{M+2}$
and $a_{s,i}$ with $s+i=M-1$ in terms of
$e_{M}$ and the coefficients of lower order. 
Then $e_{M}$ is determined by
the normalization equation, but only at order $M+2$. In this manner,
all the coefficients in \figexp\ are determined unambiguously.

The $1/\mu$ expansion described above
can be easily carried out to very high orders. 
Let us present here just the first few terms of this expansion,
\eqn\firstfew{
\left\{\eqalign{
\mu^2&=\mu_{\rm r}^2\biggl(1-{8\over \mu_{\rm r}^4}-{64\over
\mu_{\rm r}^8}+ {256\over \mu_{\rm r}^{12}}-{12800\over \mu_{\rm r}^{16}}
\biggr)+{\cal O}\biggl({1\over \mu_{\rm r}^{18}}\biggr)\cr
g_{\rm cr}&={\mu_{\rm r}^4\over 4}\biggl(1-{96\over \mu_{\rm r}^8}+{1024\over
\mu_{\rm r}^{12}}-{30720\over \mu_{\rm r}^{16}}
+{819200\over\mu_{\rm r}^{20}} 
\biggr)+{\cal O}\biggl({1\over \mu_{\rm r}^{20}}\biggr)\cr
P(z)&=1+{2560\over \mu_{\rm r}^{12}}-{128z^2\over\mu_{\rm r}^{10}} \biggl(
5+{252\over\mu_{\rm r}^4}\biggr)+{5376 z^4 \over \mu_{\rm r}^{12}}+
{\cal O}\biggl({1\over \mu_{\rm r}^{14}}\biggr)\cr
}\right.}
The information about the critical properties of the theory is encoded in the 
value of $P(z=0)$. Indeed, the critical eigenvalue density $\rho(z)$ 
behaves at small $z$ as $\pi\rho(z)\approx a(\mu^2) z^2$ with
\eqn\aaeq{\eqalign{
a(\mu^2)&={\mu_{\rm r}^3\over 2\sqrt{2}}P(z=0)\cr
&={\mu_{\rm r}^3\over 2\sqrt{2}}\biggl(
1+ {1\over \mu_{\rm r}^4} \sum_{i=1}^{M} 
{e_i \over \mu_{\rm r}^{2i}}\biggr).\cr}}
If $a(\mu^2)$ is finite and nonzero, the critical index $\delta$ equals
$\delta=2$. On the other hand, a vanishing $a(\mu^2)$ would indicate a 
higher order multicritical point. Also, one could encounter a singularity
where $a(\mu^2)$ goes to infinity. This would mean that for small $x$
the density $\rho(x)$ vanishes slower than $x^2$ and the asymptotics 
$\rho(x)\propto x^2$ does not apply. 
In fact, we do expect such a singularity to occur at the Kosterlitz--Thouless 
point $\epsilon=1$ or, equivalently, $\mu^2=2$.  At that point \REFfirstpaper\
\eqn\kostsing{
\rho(x)\propto{|x|\over\log[1/(\lambda |x|)]}}
which is indeed more singular than $\rho(x)\propto x^2$.

\ifig\numericsfigure{
The plots of $a(\mu^2)$ obtained from the strong coupling expansion with
$M=8, 10, 12$ and $14$ (counting from lower left to upper right.) 
It is seen that $a(\mu^2)$ does not vanish and
exhibits a singularity around $\mu^2=2$.}
{\epsfysize3.5in\epsfbox{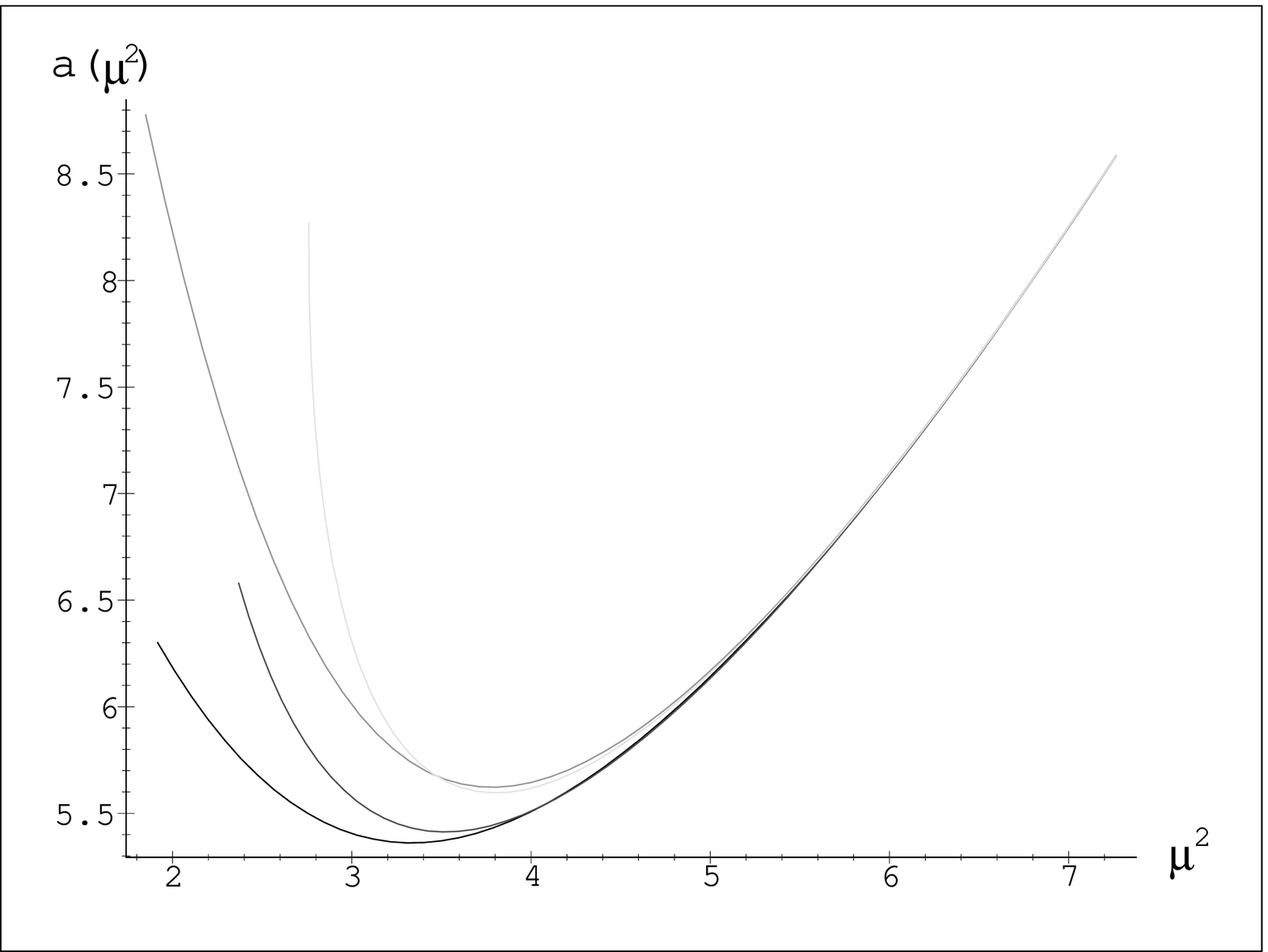}}

Therefore, the coefficient $a(\mu^2)$ should blow up at $\mu^2=2$.
In fig.~5 we plot $a(\mu^2)$ computed from the strong coupling expansion 
up to ${\cal O}(\mu^{-28})$. These plots show that $a(\mu^2)$ certainly
does not vanish. Therefore, no higher order multicriticalities occur for
$\epsilon>1$. Furthermore, $a(\mu^2)$ does blow up in the region around 
$\mu^2=2$. The Pade approximants reveal only one singularity in that
region which we can therefore associate with the Kosterlitz--Thouless 
phase transition. Of course, to determine the critical value of $\mu^2$ 
numerically is very hard, as the strong coupling expansion becomes unreliable
when  $\mu^2$ gets less than $\sim 3$. However, when combined 
with the theoretical results found in section~3, the numerical data 
become quite conclusive.  Indeed, the universality arguments suggest 
that the difference between the ``transcendental'' potentials of
section~3 and the pure quartic potential could only lead to a higher
order multicritical point in the quartic case. But such points are reliably 
excluded by the data from the strong coupling expansion. At the same time,
the exact position of the Kosterlitz--Thouless point can be derived
from equation \aone\ independently of any numerical simulations.

To summarize, the strong coupling expansion, taken together with the exact 
results, indicates that the infinite chain model with a pure quartic
potential exhibits the critical index $\delta=2$ for any $\epsilon>1$.

\newsec{Conclusions}

In this paper we have explored the $\epsilon>1$ regime of the infinite 
random matrix chain. We have calculated the critical exponents and found
that, for $\epsilon>1$, they are the same as the critical exponents in
one-matrix models. As a consequence, the critical behavior of the
strongly coupled two-dimensional $O(2)$ model on a random planar lattice
turns out to coincide with the critical behavior of pure two-dimensional
gravity.

From the technical viewpoint, we found that most properties of the chain
model are encoded in the midway density $\rho_{1/2}(x)$ which enters the
picture 
through the hydrodynamic representation. We have seen that the changes
in the branch structure of $\rho_{1/2}$ govern the Kosterlitz--Thouless
phase transition and explain the change in critical exponents.

Finally, we constructed the character expansion for the matrix chain.
Rather remarkably, in the character expansion language the mechanisms of the
Kosterlitz--Thouless phase transition and the Douglas--Kazakov-type
transitions are precisely the same. We derived an exact relation 
between the ``Young tableau density'' of the representation
dominating the character sum at large $N$ and the hydrodynamic density
$\rho_{1/2}$.

There are, however, a number of problems that remained open.
First of all, it would be desirable to engineer an exact solution
reproducing the critical behavior of the matrix chain
on both sides of $\epsilon=1$. According to 
\kostsing, the requisite midway density $\rho_{1/2}$ would have to contain a 
logarithmic singularity and, for $\epsilon>1$, have the two-branch analytic 
structure. We are confident that such a solution exists and can be constructed
with our methods.

Furthermore, it would be very interesting to interpret our results in terms 
of another theory describing the two-dimensional $O(2)$ model---the matrix
quantum mechanics on a circle. There, too, the representation structure 
changes at the Kosterlitz--Thouless point, although in a somewhat 
different way. Since the chain model and the compactified matrix
quantum mechanics are dual to each other, there might exist a correspondence
between the large $N$ limits of their character expansions.

In this respect, the matrix theory on a circle resembles the 
instanton description of QCD$_2$. Indeed, two-dimensional QCD
on a sphere has two dual descriptions. One of them is the character 
expansion given by the Migdal--Rusakov 
formula \REFdouglaskazakov\REFmigdalrusakov. 
The other represents
the partition function of QCD$_2$ as a sum over all instanton sectors,
and can be obtained from the character expansion by Poisson 
resummation \REFgrossinstantons.

At large $N$ the representation dominating in the character expansion formula
has ${\cal O}(N)$ boxes and, depending on the phase, may or may not have a gap.
The Young tableaus corresponding to such representations are shown in fig.~1.
On the other hand, for $N\to\infty$ the instanton expansion is also 
dominated by a single sector. The $N$ charges parametrizing the
dominant instanton 
all strictly equal zero 
at weak coupling but become nonzero integers of order $N$ at 
strong coupling \REFgrossinstantons\REFgrosscylinder.

This is very similar to what happens in the matrix quantum mechanics on a 
circle. There the vortexless phase is dominated by the trivial (singlet)
representation, with zero boxes in each Young tableau 
row \REFgrossklebanov\REFboulatovkazakov.
In the other phase, where the vortices are significant, the dominant 
representation has ${\cal O}(N)$ boxes. We see that
the row lengths of the dominant Young tableau in the matrix quantum mechanics
behave the same way as the instanton charges in QCD$_2$.

All of this suggests that, by analogy with QCD$_2$, a simple Poisson
resummation of the character expansion formula \chexpan\
could tell us a lot about compactified matrix quantum mechanics.
If true, such analysis
could be extremely useful, as the matrix quantum mechanics
on a circle so far appeared intractable by any known computational method.

\bigskip
It is our pleasure to thank D. Gross, V. Kazakov and A.A. Migdal for their
interest, support and for helpful advice. 

This work is supported in part by funds provided by the U.S. Department of 
Energy under cooperative research agreement DE-FC02-94ER40818 and the
Swiss National Science Foundation.

\listrefs

\end